\documentclass[epj]{svjour}
\usepackage{graphics}

\usepackage{lineno}

\begin{document}

\title{Test of the CLAS12 RICH large scale prototype in the direct proximity focusing configuration}
\author{S. Anefalos Pereira\inst{3} \and 
N. Baltzell\inst{1} \and 
L. Barion\inst{4} \and
F. Benmokhtar\inst{2} \and
W. Brooks\inst{11} \and
E. Cisbani\inst{6} \and
M. Contalbrigo\inst{4} \and
A. El Alaoui\inst{1}\thanks{\emph{Present address:} Universidad Tecnica Federico Santa Maria, Valparaiso, Chile} \and
K. Hafidi\inst{1} \and
M. Hoek\inst{10}\thanks{\emph{Present address:} Institut f{\"u}r Kernphysik, Johannes Gutenberg-Universit{\"a}t Mainz, Johann-Joachim-Becher-Weg 45, D 55128 Mainz, Germany} \and
V. Kubarovsky\inst{8} \and
L. Lagamba\inst{7} \and
V. Lucherini\inst{3} \and
R. Malaguti\inst{4} \and
M. Mirazita\inst{3}\thanks{\emph{Corresponding author:} marco.mirazita@lnf.infn.it} \and
R.A. Montgomery\inst{3,10} \and
A. Movsisyan\inst{4} \and
P. Musico\inst{5} \and
A. Orlandi\inst{3} \and
D. Orecchini\inst{3} \and
L.L. Pappalardo\inst{4} \and
R. Perrino\inst{7} \and
J. Phillips\inst{10} \and
S. Pisano\inst{3} \and
P. Rossi\inst{3,8} \and
S. Squerzanti\inst{4} \and
S. Tomassini\inst{3} \and
M. Turisini\inst{4,11} \and
A. Viticchi\`{e}\inst{3}
}                     
%
%
\institute{Argonne National Laboratory, Physics Division, 9700 S. Cass Ave, Argonne IL, 60439, USA \and 
Duquesne University, Department of Physics, 317 Fisher Hall, Pittsburgh, PA 15282, USA \and 
INFN Laboratori Nazionali di Frascati, Via Enrico Fermi 40, 00044 Frascati, Italy \and 
INFN Sezione di Ferrara, Polo Scientifico e Tecnologico. Edificio C. Via Saragat 1, I-44122 Ferrara, Italy \and 
INFN Sezione Genova, Via Dodecaneso 33, 16146, Genova, Italy \and
INFN Sezione di Roma, gruppo Sanit\`{a} and Istituto Superiore di Sanit\`{a}, I-00161 Rome, Italy \and
INFN Sezione di Bari, Via E. Orabona n.4, I-70124 Bari, Italy \and
Jefferson Laboratory, Thomas Jefferson National Accelerator Facility, 12000 Jefferson Avenue, Newport News, VA 23606, USA \and
Universit\`{a} degli Studi di Ferrara, Via Saragat 1, I-44122 Ferrara, Italy \and
University of Glasgow - School of Physics and Astronomy, Kelvin Building, University Avenue, Glasgow G12 8QQ, Scotland, UK \and
Universidad Tecnica Federico Santa Maria, Valparaiso, Chile}
\date{Received: date / Revised version: date}
%
\abstract{
A large area ring-imaging Cherenkov detector has been designed to provide clean hadron identification capability in the momentum range from 3 GeV/c up to 8 GeV/c for the CLAS12 experiment at the upgraded 12 GeV continuous electron beam accelerator facility of Jefferson Laboratory. 
The adopted solution foresees a novel hybrid optics design based on aerogel radiator, composite mirrors and highly packed and highly segmented photon detectors.
Cherenkov light will either be imaged directly (forward tracks) or after two mirror reflections (large angle tracks).
We report here the results of the tests of a large scale prototype of the RICH detector performed with the hadron beam of the CERN T9 experimental hall for the direct detection configuration.
The tests demonstrated that the proposed design provides the required pion-to-kaon rejection factor of 1:500 in the whole momentum range.
} 
\titlerunning{Test of the CLAS12 RICH prototype}
\maketitle
%

\section{Introduction}
\label{intro}

The Jefferson Laboratory (JLab, USA) main facility is currently completing a major upgrade program that already doubled the energy of the electron beam (from 6 GeV to 12 GeV).
This program will lead in the near future to an increased luminosity, the enhancement of the detector systems in the three existing Halls A, B and C and the construction of a new experimental Hall D.
The Hall B will house the new CLAS12 large acceptance multi-purpose spectrometer, that will operate with a highly polarized electron beam with energy up to 11 GeV and a luminosity as high as $10^{35}$cm$^{-2}$s$^{-1}$ impinging on unpolarized as well as polarized hydrogen and deuterium targets, thus providing unique conditions for the study of electron-nucleon scattering \cite{CLAS12}.

The CLAS12 experimental program \cite{CLAS12_phys} covers many topics in hadron physics.
Particular attention is devoted to the 3D imaging of the nucleon through the study of generalized and transverse momentum dependent parton distributions (GPDs and TMDs) in the poorly explored valence region (high Bjorken x).
A number of approved experiments \cite{PSHP} require an efficient hadron identification in the 3-8 GeV/c momentum range in order to allow the flavor separation. 
As the flux of pions is one order of magnitude larger than kaons, a pion rejection factor of about 1 : 500 is required to limit the pion contamination in the kaon sample to a few percent level. 

The standard particle identification (PID) is performed in CLAS12 by using a time-of-flight (TOF) system and two Cherenkov gas detectors with high (HTCC) and low (LTCC) thresholds.
The TOF is able to provide full hadron PID up to about 3 GeV/c, while the two Cherenkov detectors reach the required pion rejection factor only at the edge of the available phase space (hadron momenta above 7 GeV/c) and are not able to distinguish kaons from protons.
In order to achieve the neccessary PID requirements in the full momentum range, a modular Ring Imaging Cherenkov (RICH) detector has been proposed to replace at least two of the six azimuthal sectors of the LTCC.
Each RICH sector will have a projective geometry, with a depth of about 1.2 m and almost $5$ m$^2$ entrance window area.

Based on simulation studies \cite{RICH_sim1,RICH_sim2}, the best configuration for the detector is a non-conventional proximity-focusing design, with a wall of aerogel radiator tiles, an array of visible light photon detectors and a mirror system.
The latter is essential to reduce to about 1 m$^2$ per RICH sector the area covered by the photon detectors, thus minimizing costs and the material-budget impact on the downstream CLAS12 detectors (TOF and Calorimeters). 

\begin{figure}
\begin{center}
\resizebox{0.4\textwidth}{!}{%
  \includegraphics{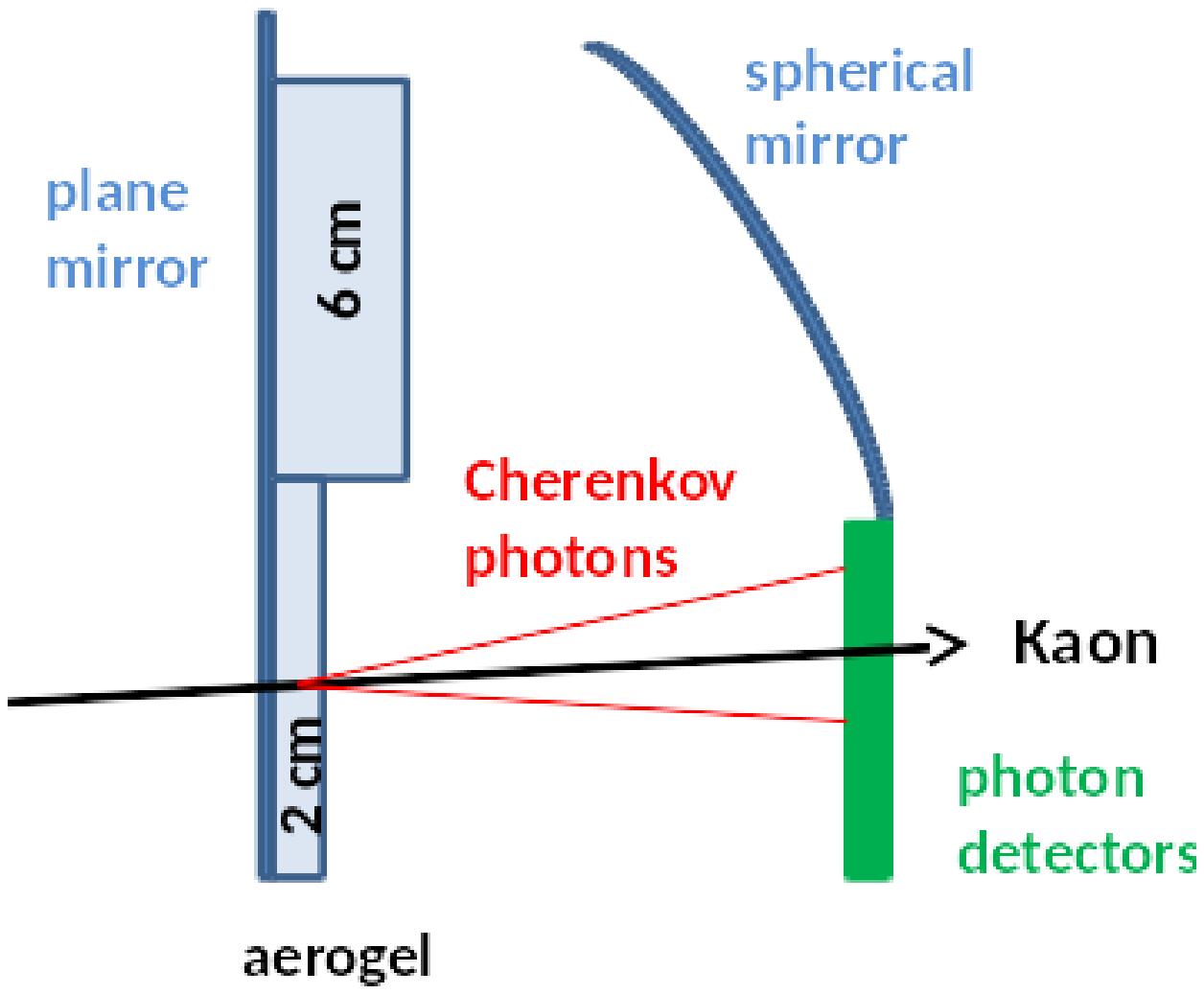}
}
\resizebox{0.4\textwidth}{!}{%
  \includegraphics{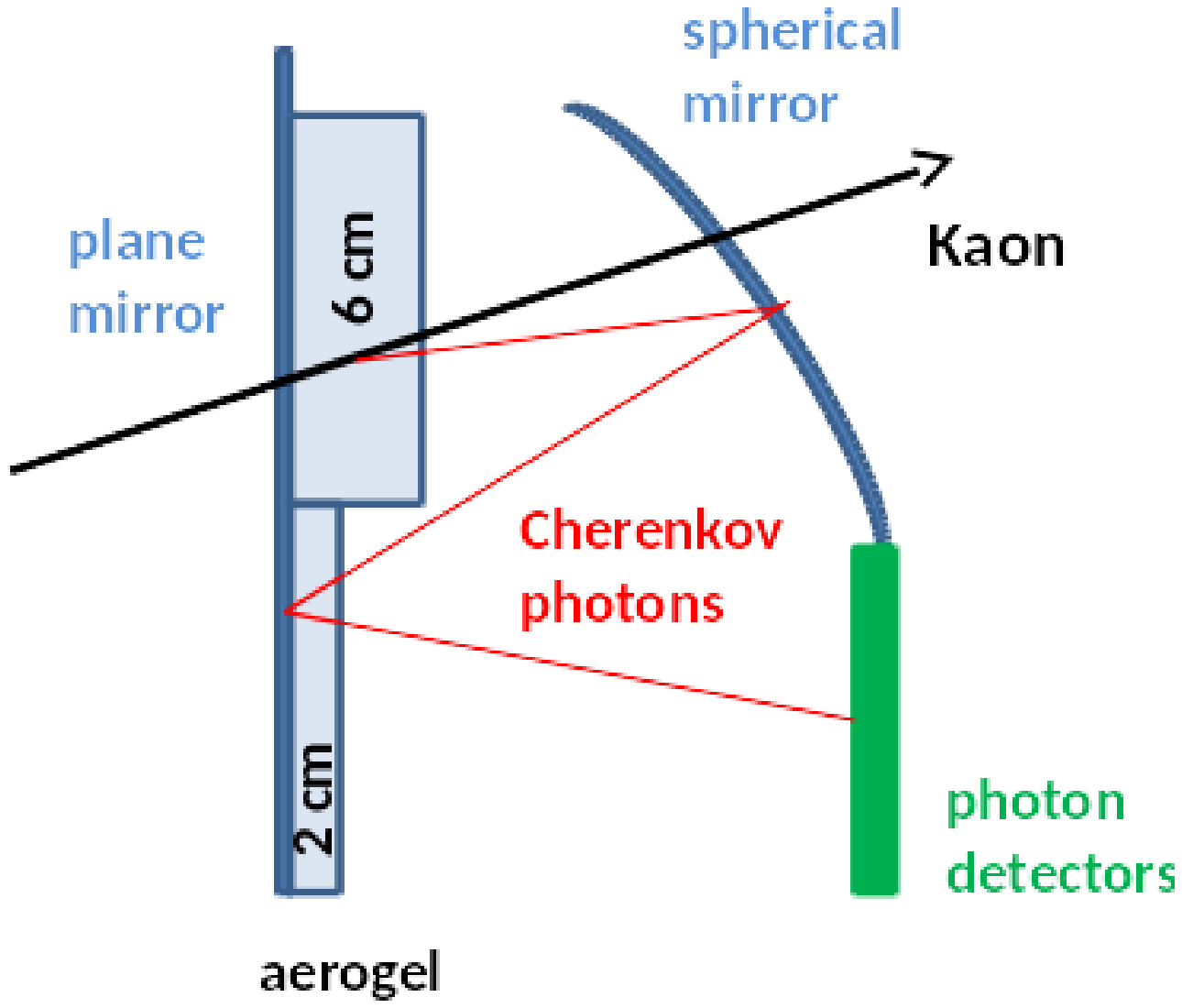}
}
\caption{The conceptual design of the RICH detector for forward (top plot) and large angle (bottom plot) particles.
}
\label{fig:RichConcept}
\end{center}
\end{figure}

The concept of this RICH is illustrated in Fig. \ref{fig:RichConcept}. 
For forward scattered particles ($\theta < 12^o$) with momenta in the range between 3 and 8 GeV/c, the Cherenkov light will be directly detected by the photon detector array (Fig. \ref{fig:RichConcept} top). 
For particles with larger incident angles ($12^o < \theta < 35^o$), with momenta between 3 and 6 GeV/c, the Cherenkov light will be double-reflected by a spherical and a planar mirror and focused onto the photon detector array after two passages through the lower section of the aerogel wall (Fig. \ref{fig:RichConcept} bottom). 
This double pass imposes the use of aerogel with reduced thickness (2 cm) and with relatively high refractive index ($n$=1.05) in the lower section of the detector, to minimize the photon yield loss.
The photon yield losses will be compensated by the use of a thicker (6 cm) aerogel in the upper section.

A variety of laboratory and beam tests have been performed in order to validate this innovative configuration.
Laboratory tests have been done to study the single RICH component performances, in particular the optical properties of aerogel produced by various companies worldwide and the performaces of different high-granularity photon detectors in the Single Photo-Electron (SPE) regime. 
In addition, a campaign of beam tests has been pursued using prototypes of increasing scale and complexity, performed at the Frascati Beam Test Facility \cite{BTF} and at the CERN T9 beam line \cite{CERN_T9}.
In the following, we report the results of the final tests performed at CERN for the direct light configuration, including several tests performed on the main components, whose performance was challenged in the limit of high momenta.
The prototype used in these tests was designed in such a way that it reproduces the direct light configuration of the CLAS12 RICH.

For the reflected light configuration, the prototype was designed following the same concept of the CLAS12 RICH (i.e. the use of spherical and planar mirrors and the double pass through the aerogel), although geometry limitations imposed a reduced light path length, not optimized for real hadron separation. 
Preliminary results for the reflected light configuration have been already published \cite{Cont2014}, the final results will be reported in a future paper.

\section{The experimental setup}

\subsection{The CERN T9 beam line}
\label{sect:T9line}

The setup for the tests was installed at the T9 experimental area located in the East Area of the PS/SPS complex at CERN \cite{CERN_T9}.
The primary proton beam of the PS is sent to a target to produce a secondary beam.
Different targets, magnets and collimators allowed to select the composition (electrons, muons or hadrons), the charge and the momentum \cite{CERN_T9_optics} of the secondary beam.

The tests were performed using the settings for negative charge hadrons with the highest momentum possible, allowing to cover the region between 6 and 8 GeV/c, the most critical one for the CLAS12 RICH.
The relative population of $\pi^- : K^- : \bar{p}$ was approximately $160 : 5 : 1$, basically independent from the momentum in the range between 6 and 8 GeV/c.
A number of dipole and quadrupole magnets with predetermined settings allowed to focus the beam  at different distances from the beam pipe exit windows.
The choice was to have the smallest beam spot as close as possible to the RICH photodetector plane, at 10 meters from the exit window of the beam line.

The time structure of the beam was determined by the number of spills (from 1 to 3) extracted from the primary PS proton beam.
Each spill had a duration of 400 ms and the period of the PS operation was 40 s, corresponding to an overall duty cycle of the order of few percent.

The T9 equipment included a CO$_2$ gas Cherenkov counter used to separate pions from heavier hadrons by adjusting the gas pressure so that only pions were above threshold.

\subsection{The RICH prototype}
\label{sec:prototype}

\begin{figure}
\begin{center}
\resizebox{0.45\textwidth}{!}{%
  \includegraphics{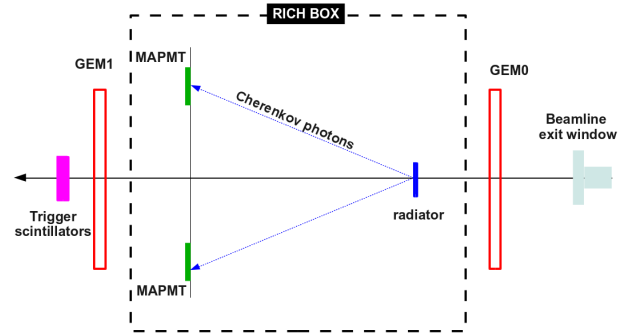}
}
\resizebox{0.45\textwidth}{!}{%
  \includegraphics{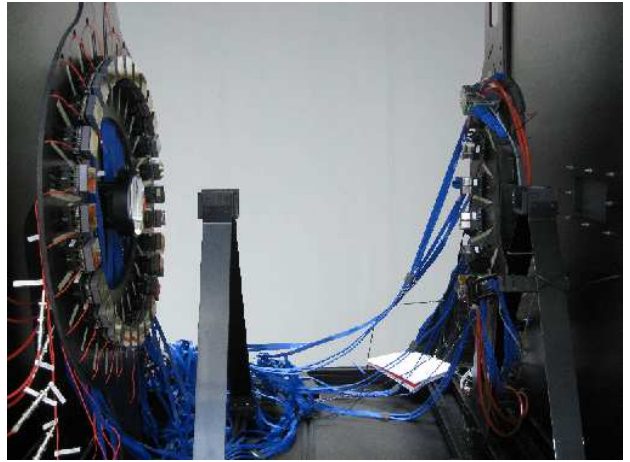}
}
\caption{Top plot: a schematic drawing of the setup in the T9 experimental hall.
Bottom plot: picture of the RICH box installed at the CERN-T9 beam line; the beam enters from the right; the photon detector array is on the left of the picture, while the support for the aerogel tiles is on the right.}
\label{fig:Setup_direct}
\end{center}
\end{figure}

The RICH prototype has been designed and built at the INFN Laboratory of Frascati.
It comprises a large (approximately $1.6 \times 1.8 \times 1.6$ m$^3$) light-tight box and internal modular supports holding the various components, that may be adjusted depending on the measurement conditions.
The proximity gap length was about 1 m, close to the CLAS12 conditions.
A schematic drawing of the setup and a picture of the RICH box during the installation are shown in Fig. \ref{fig:Setup_direct}.

\subsubsection{The radiator} 

An excellent radiator for RICH hadron identification in the few GeV momentum range is silica aerogel, an amorphous solid network of $\rm{ Si O_2}$ nanocrystals with a very low macroscopic density and a refractive index in between gases and liquids. 
It has been used as radiator material for RICH detectors in several particle physics experiments \cite{Aerogel_old1,Aerogel_old2,Aerogel_old3,Aerogel_old4} and is planned for future use \cite{Aerogel_new}.

A systematic characterization has been carried out on a variety of aerogel samples from different manufacturers. 
The aerogel from the Budker and Boreskov Catalysis Institutes of Novosibirsk \cite{BINP1,BINP2}, that combines high transparency with a considerable geometrical (area and thickness) flexibility and mass production capabilities, was chosen.
Noteworthy, during the prototyping, the production technique and the resulting quality have significantly improved in time, reaching a clarity parameter as low as 0.0050 $\mu$m$^4$ cm$^{-1}$ for a n = 1.05 refractive index at the reference wavelength $\lambda_0=400$ nm (blue light).

Precise measurements of the aerogel transmittance have been performed, using a Lambda~650~S~PerkinElmer spectrophotometer, on a broad range of light wavelengths between 200 and 900 nm \cite{AerogelTests}. 
Typical transparency values in the blue light region are close to $70 \%$ for the most recent production of tiles with refractive index of 1.05 and 20 mm thickness, while lower values, close to $60 \%$, have been measured for the older production of aerogel with refractive index of 1.04 and 1.06.

The Cherenkov light emission spectrum per unit radiator thickness $x$ and wavelength $\lambda$ can be written as
\begin{equation}
\label{eq:CerSpectrum}
\frac{d^2N_{C}}{dxd\lambda} = \frac{2 \pi \alpha}{\lambda^2} \sin^2(\eta_C)
\end{equation}
where $\eta_C$ is the Cherenkov emission angle.
The spectrum has a broad distribution and it is rapidly increasing when the wavelength changes from the visible light toward the UV region.
Thus, it is important to keep under control the chromatic dispersion, i.e. the dependence of the refractive index on the light wavelength, as it constitutes one of the largest contributions to the Cherenkov angle resolution.
One of the standard techniques for its measurement is the so-called prism method in conjunction with the Snell-Descartes formula, but experimental measurements exist in literature only for refractive index up to $n=1.03$ \cite{SnellDescartes}.

In our prototype, we used aerogel tiles with an area of approximately $58 \times 58$ mm$^2$, 20 mm thickness and refractive indices $n=$ 1.04, 1.05, 1.06.
The Novosibirsk aerogel is hydrophilic, i.e. tends to absorb water molecules from the air, resulting in a degraded transparency. 
Thus, it was important to monitor the transparency of each individual aerogel tile during the data taking and eventually correct the data for the degraded transparency. 
For this reason, a strict protocol to store and handle the aerogel was established.
In addition, before and after the installation, the relative transparency of each tile was monitored using a simple setup made by a blue light laser ($\lambda=405$ nm), a photodiode and a multimeter. 
The transmittance was then evaluated as the ratio between the multimeter readout with and without the aerogel.
This method has the advantage of being very fast (a few minutes for each set of measurements) but provides only relative values of transparency. 
The results showed only marginal degradation in the transparency between the measurements performed before and after the usage of the tiles, thus no corrections have been applied to the data.

\subsubsection{The photodetectors}

Dedicated simulation studies \cite{RICH_sim} have shown that the required Cherenkov angle resolution can be achieved if the pixel size of the photon detectors is less than 1 cm.

The Hamamatsu H8500 multianode photomultiplier tubes (MAPMTs) \cite{HamamatsuH8500} have been selected as a good candidate for the first module of the CLAS12 RICH, being an effective compromise between detector performance and cost.
It comprises an $8 \times 8$ array of pixels, each with dimensions $5.8 \times 5.8$ mm$^2$, into an active area of $49 \times 49$ mm$^2$  with a very high packing fraction of $\epsilon_{pack} \approx 89\%$ and a collection efficiency $\epsilon_{coll} \approx 80\%$ \cite{hamamatsu_pmt}. 
The device offers a spectral response matching the spectrum of light transmitted by the aerogel, with a quantum efficiency peaking at 400 nm, and a fast response (less than 1 ns rise time) useful to suppress the background.

Although the H8500 MAPMT is not advertised as the optimal device for single photon detection purposes, the characterizations performed in laboratory on several units in this regime \cite{MAPMT_LNF} have demonstrated that this device can achieve performances adequate for the CLAS12 RICH requirements.
The uniformity of the H8500 response has been extensively studied with a pico-second pulsed laser. 
The typical gain variations in the pixel response of each MAPMT, of the order of 1:2, can be easily compensated by the readout electronics. 
Sub-millimeter precision scans have been used to study the PMT response across the MAPMT surface and to evaluate the true active areas of the pixels \cite{MAPMT_Glasgow}. 
The fraction of SPE signal below the pedestal threshold has been reduced to less than 15\% operating the high voltage at 1040V or above, while less than 5\% crosstalk has been measured.

The RICH prototype was instrumented with 28 MAPMTs, mounted on a circular support, that can be radially moved so to intercept the Cherenkov rings produced with different opening angles depending on the aerogel refractive index.
Fourteen Hamamatsu H8500C, with normal borosilicate glass window (wavelength cut-off at about 250 nm), and fourteen Hamamatsu H8500C-03, with glass window of enhanced transparency for UV photons (wavelength cut-off at about 185 nm) were alternated along the ring.
The ring coverage varied between about $80\%$ for $n=1.04$ aerogel refractive index to about $60\%$ for $n=1.06$.

In order to increase the separation of the SPE region from the pedestal, the high voltage of the MAPMTs was set at 1075V, close to the recommended upper limit of 1100V.

\subsection{The tracking system}

Particles' trajectories were reconstructed using two $10 \times 10$ cm$^2$ triple-GEM chambers \cite{GEM}.
The GEM foils and the double layer readout strip plane were produced at CERN \cite{CERN_GEM}. 
Each chamber measures the particle hits on a $(x, y)$ plane made of $256\times256$ strips, with a nominal spatial resolution of about 100 $\mu$m.

The two chambers were placed about 4.2 m apart from each other, with the upstream one placed 3.5 m before the MAPMT plane and the downstream one 0.7 m behind the MAPMT plane.

\subsection{The trigger}

The trigger of the Data Aquisition (DAQ) system was produced by two small plastic scintillators, partially overlapping on a small spot of about $2 \times 3$ cm$^2$, placed 1 m downstream of the RICH MAPMT plane and just after the second GEM chamber.
The scintillators were readout by Hamamatsu R1450 single anode PMTs.
The output signals were amplified, discriminated and then put in coincidence with the signal coming from the beam by using standard NIM electronics.
This coincidence was then triggering the readout electronics of the various systems.

\subsection{Readout electronics}

The prototype was equipped with 28 MAPMTs with a total of 1792 channels, thus requiring highly integrated and compact readout electronics.
We used custom-made electronics based on the 64 channel MAROC3 \cite{MAROC} chip, an ASIC manifactured by the Omega Group of LAL (Paris, France) for the ATLAS experiment.

The chip embeds an individual channel preamplifier stage, with variable gain, followed by a highly configurable shaping section, and provides binary and analog charge measurements on independent lines.
The binary output is produced with an adjustable threshold discriminator array, 64-channels wide, while the analog is obtained with a sample and hold circuit and it is serially multiplexed out and digitized by an external ADC.
During this test, only the serial analog output has been used.

\begin{figure}
\begin{center}
\resizebox{0.45\textwidth}{!}{%
  \includegraphics{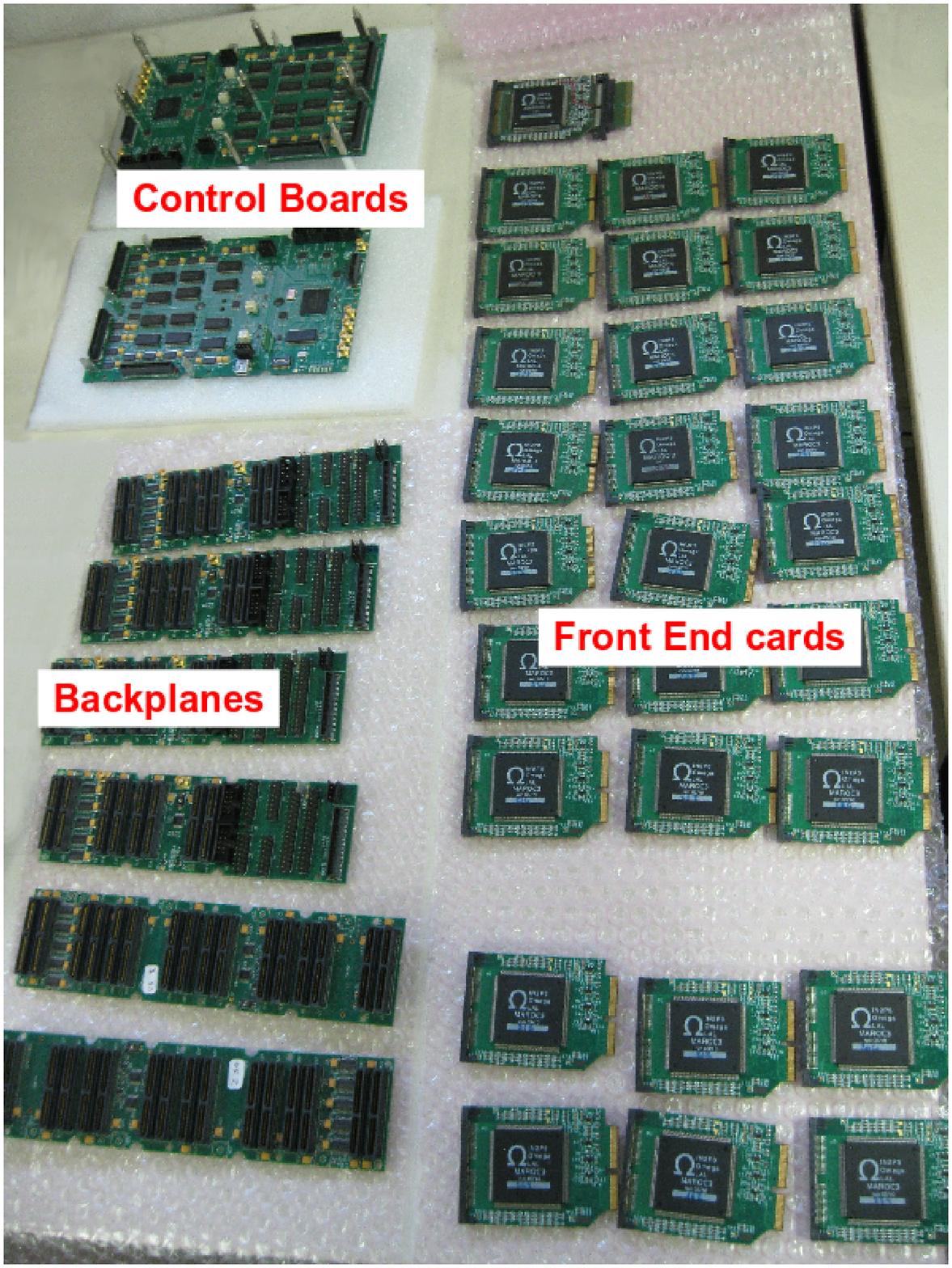}
}
\resizebox{0.45\textwidth}{!}{%
  \includegraphics{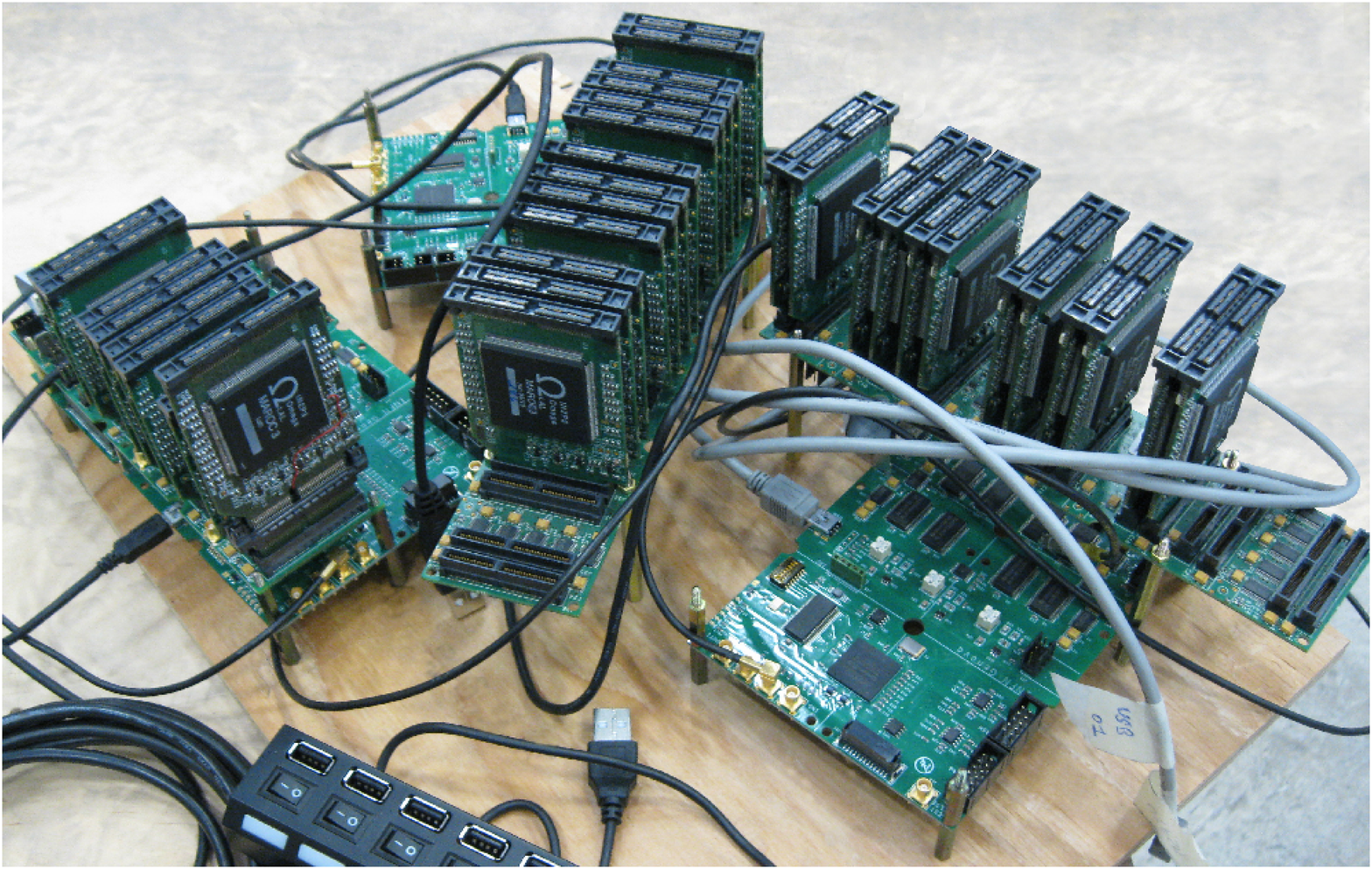}
}
\caption{Top plot: the various components of the readout electronics. Bottom plot: the assembled setup installed in the RICH prototype box.}
\label{fig:Maroc}
\end{center}
\end{figure}

Each MAROC3 chip is mounted on a front end card connected to the MAPMT through high density coaxial flat cables and equipped with an FPGA in charge of the configuration, the control and the readout of the chip.
The 28 front end cards are installed on 4 backplanes that provide electrical connectivity for power supply and data transfer.
Two control boards act as event builder and provide USB2.0 interface to the aquisition node.
The total number of available channels is 4096 in a very compact system \cite{Argentieri}.
The various components and the assembled set-up are shown in Fig. \ref{fig:Maroc}.

The GEM electronics \cite{Bellini} is based on dedicated VME modules and on the analog APV25 chip developed by Imperial College for the CMS silicon detectors \cite{APV25} and first used in a GEM detector by COMPASS. 
The APV25 is designed to be radiation tolerant (at least up to 10 Mrad), has a high channel density (128 channels/chip) and a rather fast readout (up to 40 MHz). 
Each APV25 channel consists of a preamplifier followed by a shaper and a 192 cell analog pipeline, where the shaped signal is continuously sampled at 20 or 40 MHz. 
The samples awaiting readout are flagged by external triggers, with adjustable latency. 
Multiple samples (up to 6) per channel and event will be converted and acquired by the GEM electronics. 
The acquired information permits a simple signal analysis which provides time (with resolution of about 5 ns) and charge information.

The DAQ hardware electronics was completed by a CAEN V785N ADC for the threshold Cherenkov counter readout.

The DAQ software managed the acquisition of the different systems through a CAEN V2718 VME bridge, connected to the computer via optical link.
Events for each electronic system are flagged by the DAQ with a common event number which allows the full and unambiguous reconstruction of the event.

\section{Data reconstruction}


\subsection{The gas threshold Cherenkov counter}

\begin{figure}
\begin{center}
\resizebox{0.45\textwidth}{!}{%
  \includegraphics{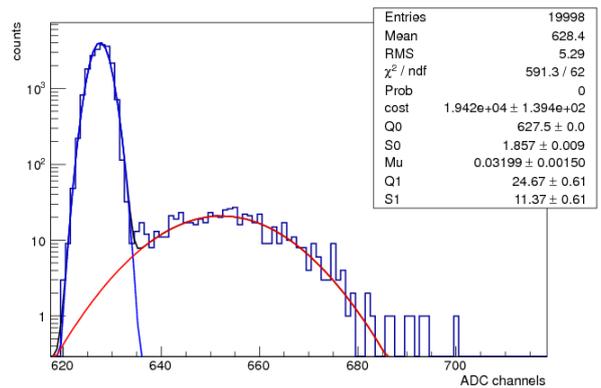}
}

\caption{Typical ADC spectrum of the threshold Cherenkov counter. The dashed red line indicates the threshold to separate pions from heavier hadrons (5 $\sigma$ above the pedestal mean).}
\label{fig:CerAdc}
\end{center}
\end{figure}

A typical spectrum of the gas threshold Cherenkov counter is shown in Fig. \ref{fig:CerAdc}, where the broad distribution is due to the pions while the narrow pedestal peak is due to hadrons below the Cherenkov emission threshold.
The cut at 150 channels (5$\sigma$ above the pedestal peak) is set to separate pions from heavier hadrons.

\subsection{The photodetectors}

The electronic system based on the MAROC3 chip allowed to individually record ADC information of all the $28 \times 64 = 1792$ pixels of the MAPMTs.
In order to minimize the SPE loss, a uniform gain $g=4$ has been set to all the MAROC3 channels.
No finer channel-to-channel equalization has been performed.

\begin{figure}
\begin{center}
\resizebox{0.5\textwidth}{!}{%
  \includegraphics{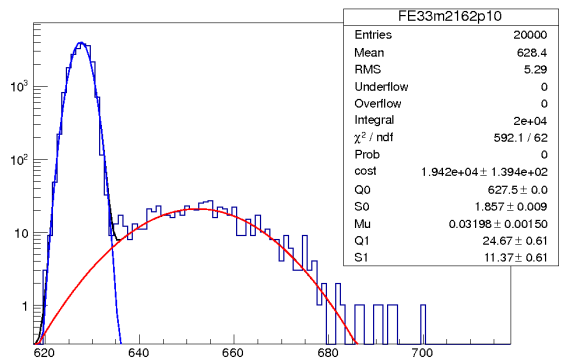}
}

\caption{Typical ADC spectrum (in logarithmic scale) of one of the MAPMT pixels registered by the MAROC3 electronics. The histogram has been fitted with the sum of two gaussian curves for the pedestal and the first photoelectron peak, shown by the curves in the plot. The fitted parameters are reported in the legend and described in the text.}
\label{fig:MarocAdc}
\end{center}
\end{figure}

A typical ADC spectrum registered from one of the channels is shown in Fig. \ref{fig:MarocAdc}: the pedestal and the SPE regions are clearly visible and well separated.
The figure shows also a fit with the sum of two gaussians, one for the pedestal and one for the SPE signal, weighted by the Poisson probability to have zero or one photoelectron, respectively.
In the legend, the values of the gaussian mean and sigma of the pedestal ($Q_0$ and $S_0$) and of the SPE peak ($Q_1$ and $S_1$) as well as the average number of photoelectrons $\mu$ and the normalization constant are reported.
A cut at $5 \sigma$ above the pedestal peak selects the Cherenkov photon hits.
The SPE efficiency has been estimated by extrapolating the SPE gaussian of the fit in Fig. \ref{fig:MarocAdc} below the pedestal cut.
We obtained $\epsilon_{\rm{SPE}} \approx 0.87$.

The ADC spectra of all the pixels have been carefully analysed in order to identify and eventually remove noisy or dead channels.
At the end of this data quality analysis, about 20 bad channels have been found, corrresponding to about $1 \%$ of the total.

\begin{figure}
\begin{center}
\resizebox{0.35\textwidth}{!}{%
  \includegraphics{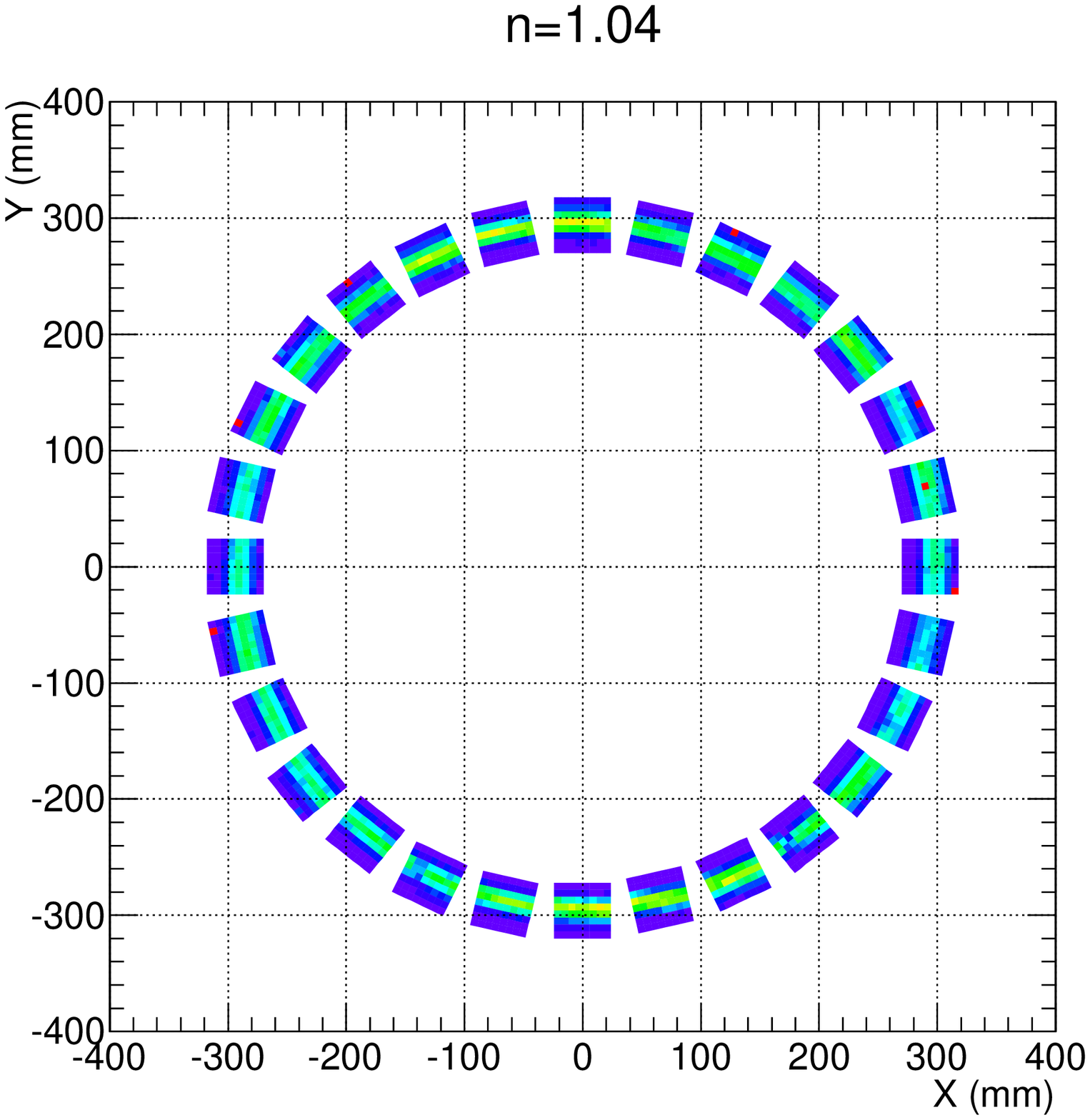}
}
\resizebox{0.35\textwidth}{!}{%
  \includegraphics{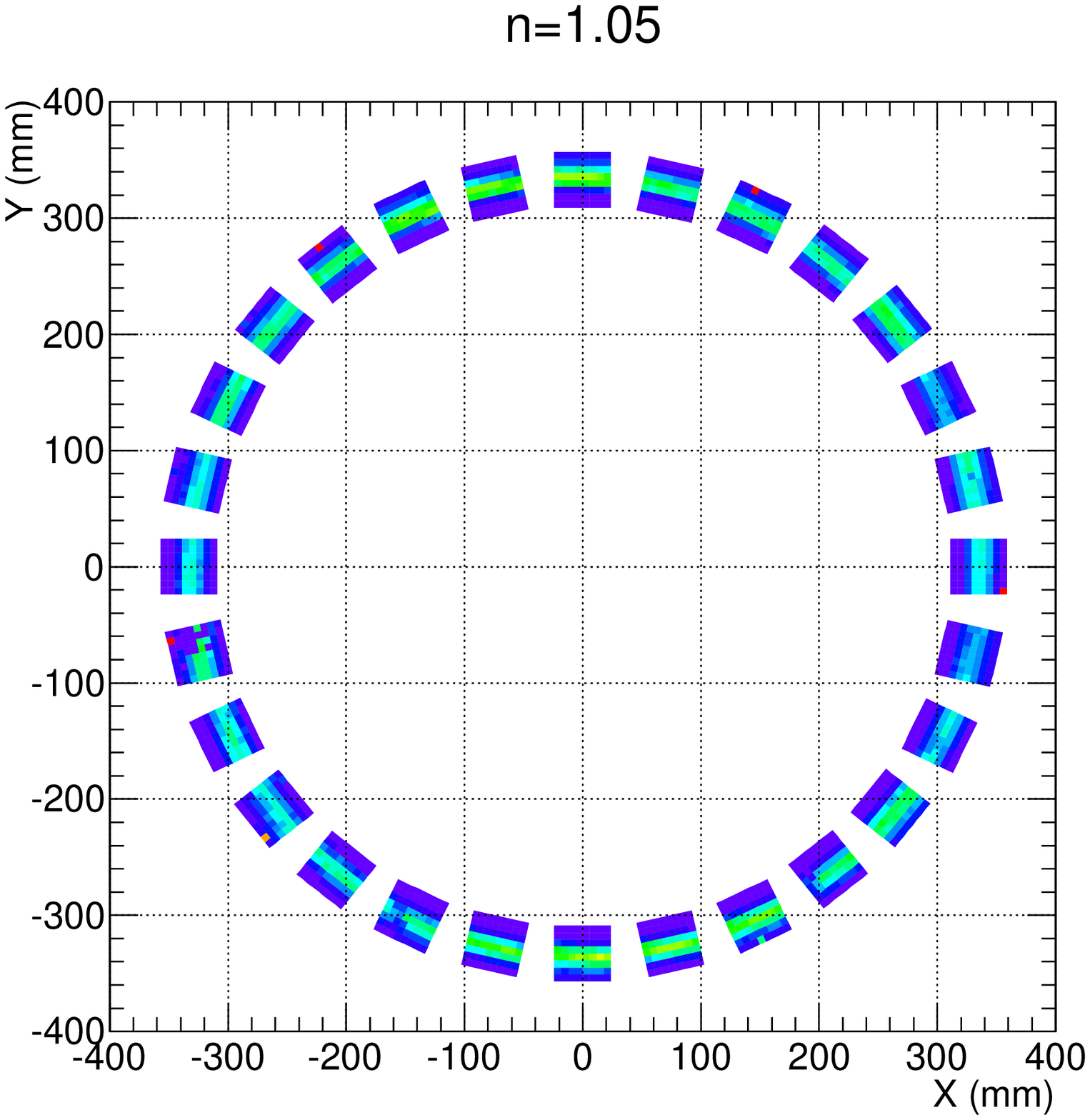}
}
\resizebox{0.35\textwidth}{!}{%
  \includegraphics{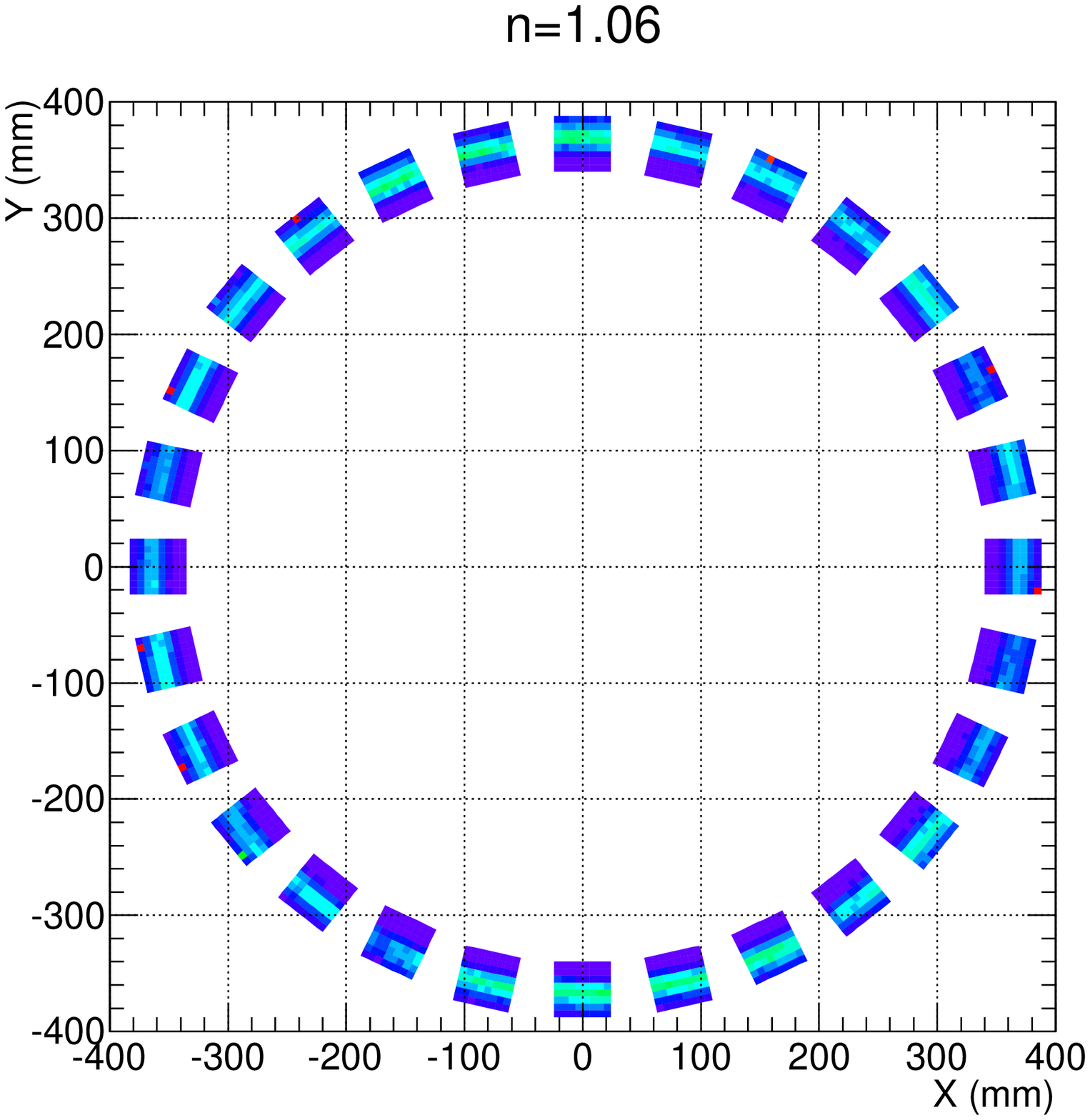}
}

\caption{Online monitoring of the Cherenkov photon hit patterns measured with aerogel of different refractive index $n$. Data with 8 GeV/c beam momentum.}
\label{fig:HitPattern}
\end{center}
\end{figure}

For each run, an event display allowed to monitor online the quality of the data.
Examples of the measured hit distributions for runs with different aerogel refractive index $n$ are shown in Fig. \ref{fig:HitPattern}.
Rings of increasing radius from $n=1.04$ to $n=1.06$ can be seen.

\subsection{The GEMs}

The GEM tracking system consists of a pair of chambers, each one with readout on the $(x, y)$ strip planes, located upstream and downstream of the RICH. 
For each strip, three samples of the charge signal, separated by 25 ns, are recorded when a trigger is issued. 
The track reconstruction proceeds through the following two steps.

\begin{figure}
\begin{center}
\resizebox{0.5\textwidth}{!}{%
  \includegraphics{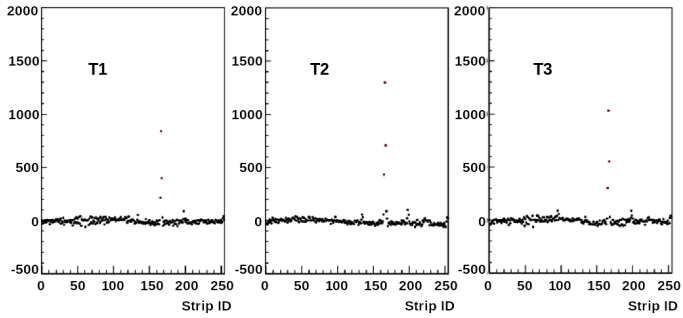}
}

\caption{Typical GEM event readout on one plane: ADC signal as a function of the strip ID, after pedestal subtraction, at the three sampling times T1, T2=T1+25 ns and T3=T1+50 ns.}
\label{fig:GemAdc}
\end{center}
\end{figure}

The first step is the identification, in a given GEM plane, of a peak at the same location (i.e. same strip) in all the three time samples.
A typical example of one event for one GEM plane is shown in Fig. \ref{fig:GemAdc}.
In this case, a good signal is easily identified on the strip 169 in all the three samples.
All the peaks exceeding 100 ADC units above the pedestal are ordered according to their amplitude and kept for further analysis.
The second step is the matching of the $x$ and $y$ peaks in one chamber to obtain a good GEM hit and finally to match a pair of good hits on the two GEM chambers to reconstruct a good hadron track.

\begin{figure}
\begin{center}
\resizebox{0.5\textwidth}{!}{%
  \includegraphics{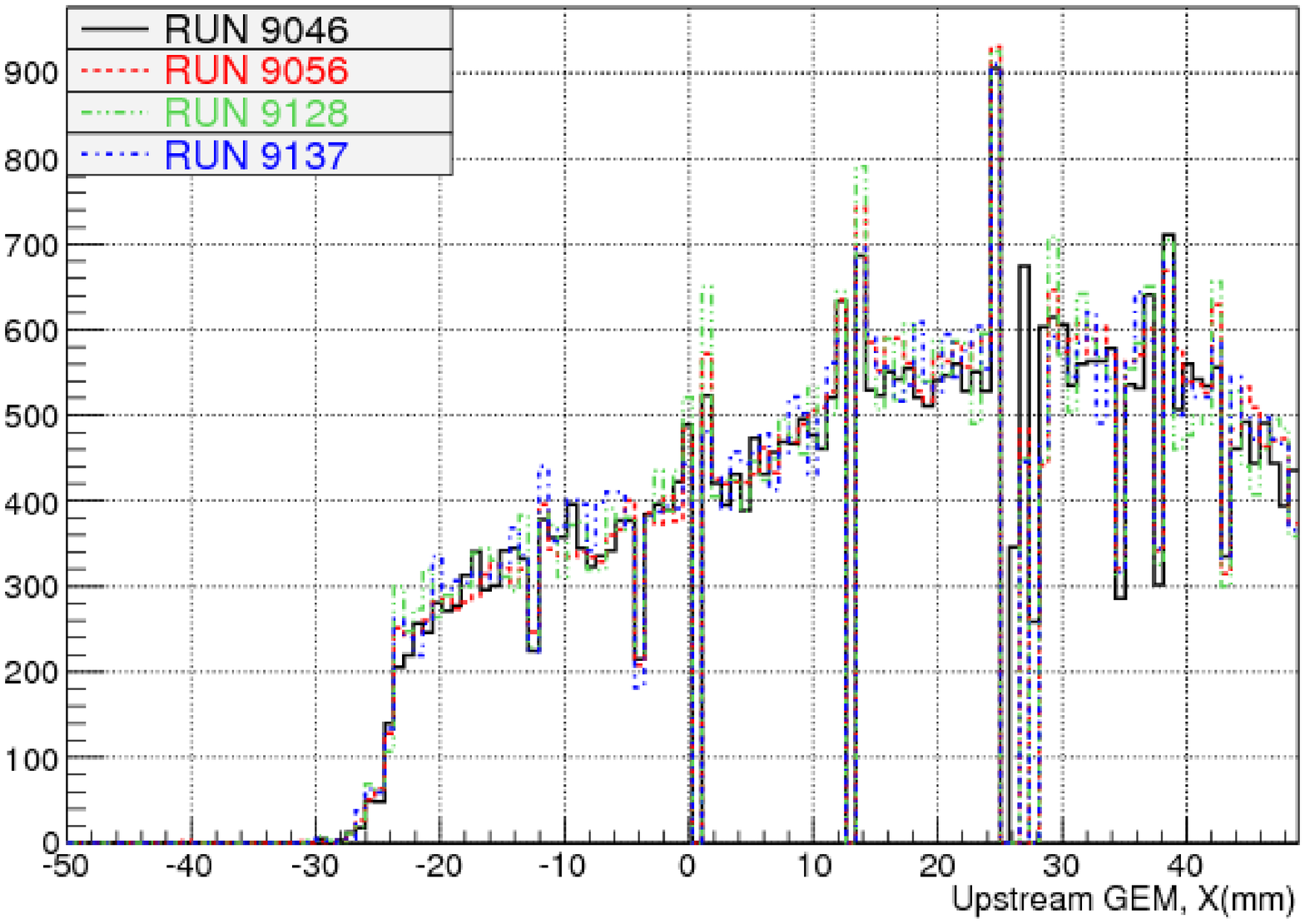}
  \includegraphics{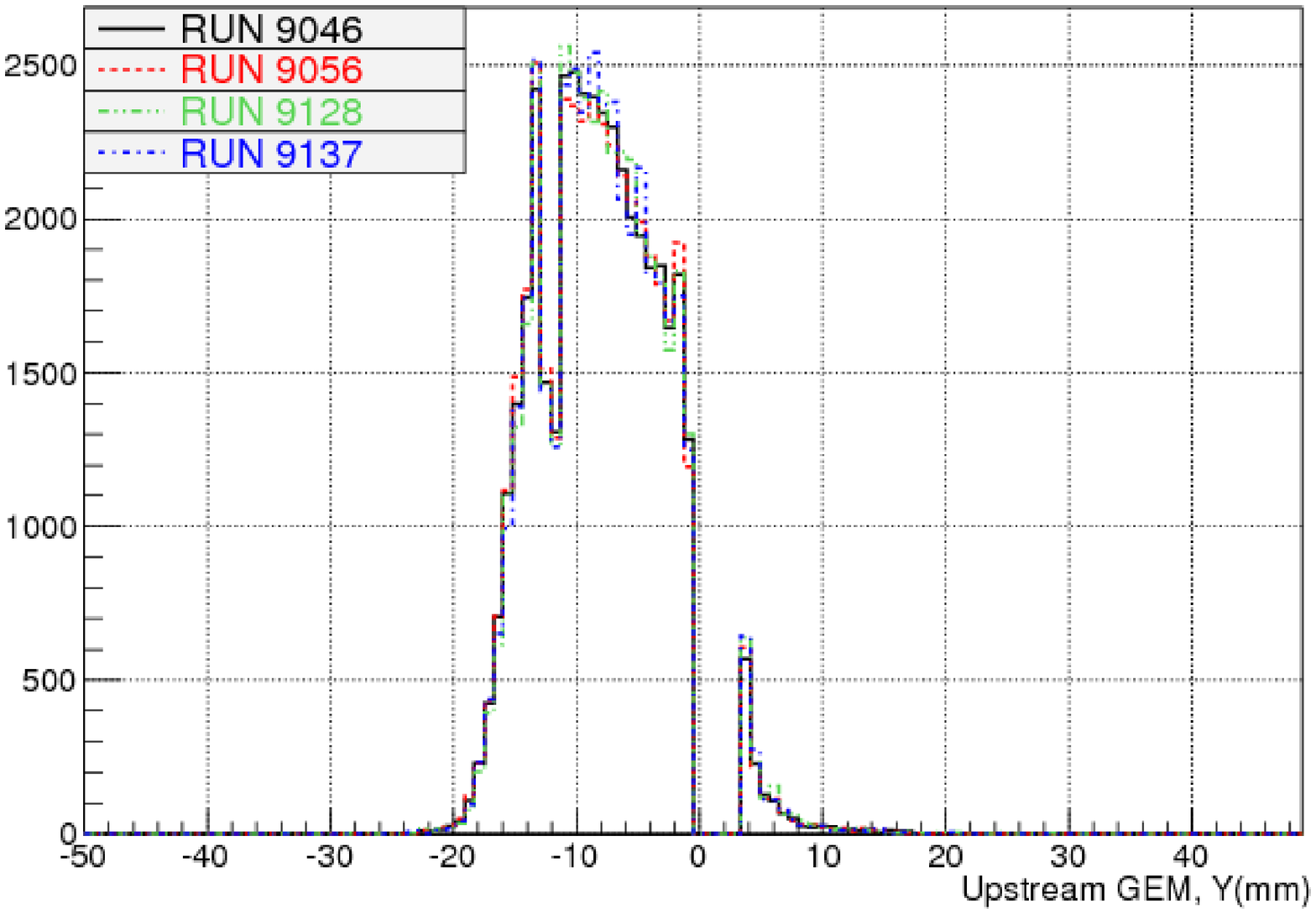}
}
\resizebox{0.5\textwidth}{!}{%
  \includegraphics{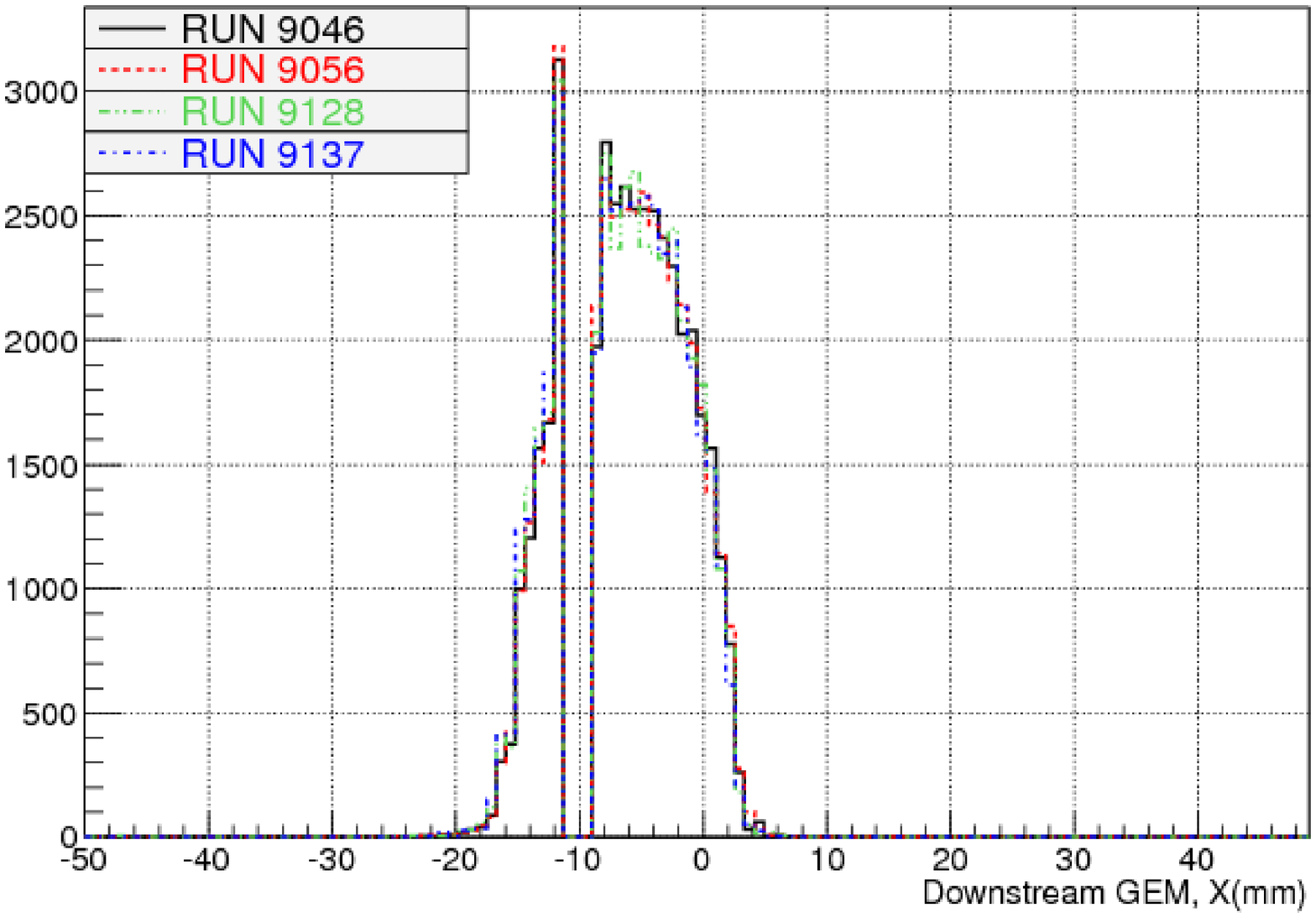}
  \includegraphics{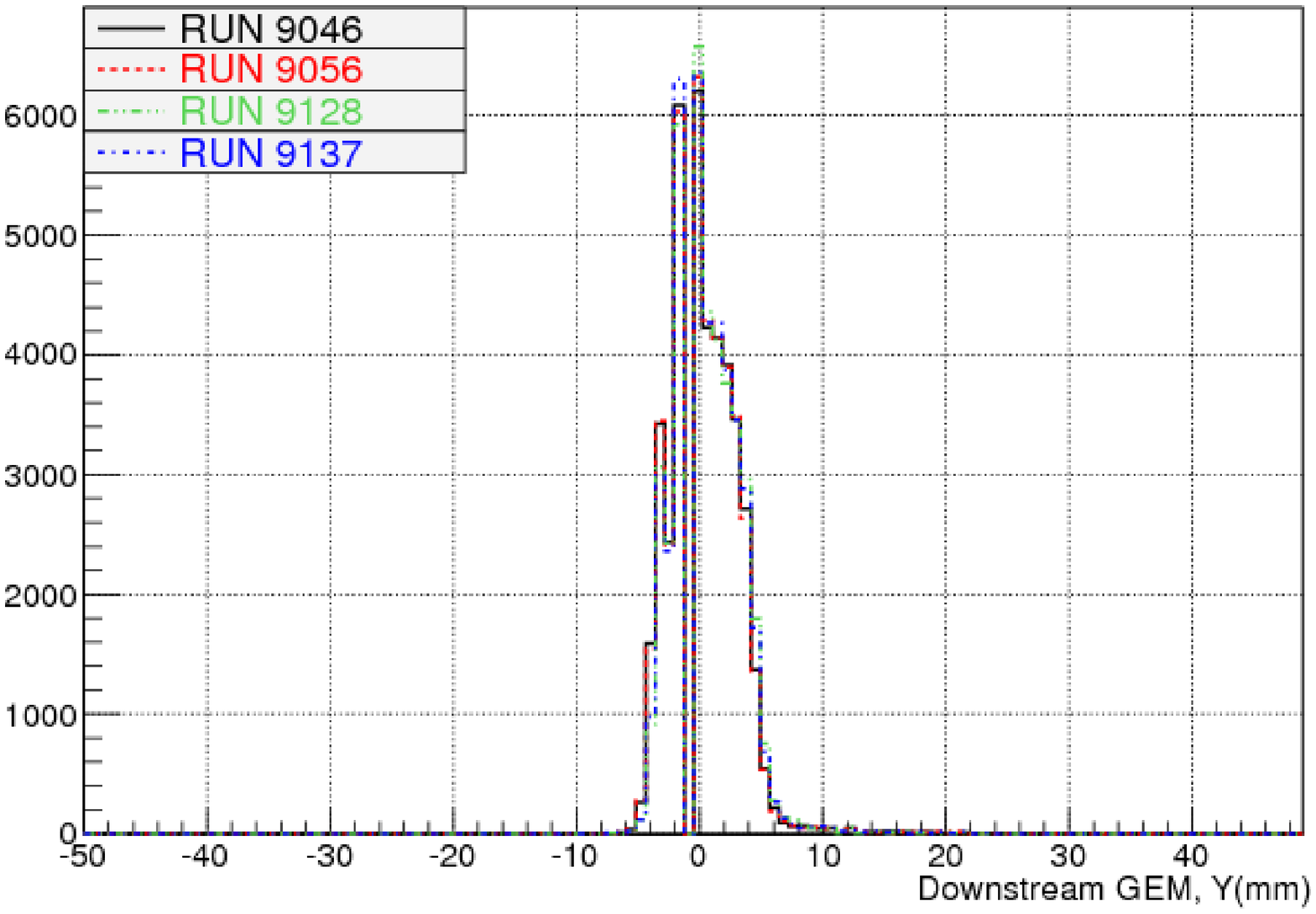}
}

\caption{Top plots: beam profiles measured on GEM0 (upstream to the RICH) in the horizontal (left plot) and vertical (right plot) directions for several runs. Bottom plots: the same for the GEM1 (downstream to the RICH). Data for 8 GeV/c momentum.}
\label{fig:GemProfiles}
\end{center}
\end{figure}

The beam profiles measured in the two GEM chambers are shown in Fig. \ref{fig:GemProfiles} for several different runs over the data taking period.
No major variations in the beam profile have been found during the test period, ensuring the stability of the beam conditions and of the tracking system.
The holes in the distributions are due to noisy strips that have been removed in the off-line data quality analysis.
The profile is wider on GEM0 (upstream GEM) than on GEM1 (downstream GEM) because of the focusing of the beam.
The horizontal width is about twice the vertical one due to a different beam collimator in $x$ and $y$.


Due to the beam structure, on average no more than one good charged track per trigger was expected.
The fraction of events with more than one reconstructed track is about $3 \%$.
This is due to either electronic noise on the GEM readout induced by the MAROC3 electronics or fake peaks produced by the reconstruction algorithm in correspondance of the removed noisy strips.
These additional hits have usually much smaller amplitude than the physical ones.
For this reason, only the track resulting from the hits with the biggest amplitude in each event is considered as a good track hit candidate.

The hit reconstruction efficiency is above $80\%$ on GEM0 and around $60\%$ on GEM1, the latter lower value due to a leak of gas that was not possible to completely fix.
Overall, the track reconstruction efficiency was about $50\%$, lower than the typical value (about $80\%$, with a $90\%$ efficiency on each chamber) expected for this type of detector.

\section{Event reconstruction}

\subsection{Alignment of the tracking system}
\label{sect:GEM_alignment}

Rough alignment of the GEM chambers was done during the installation of the setup.
A finer alignment has been obtained off-line by fitting the GEM tracks to the ring reconstruction.

A set of 4 alignment constants $({X_{0}, Y_{0}, X_{1}, Y_{1}})$ has been introduced to take into account possible displacements of the GEM on the transverse plane with respect to the beam line.
These constants have been determined by comparing the projection of the GEM tracks onto the MAPMT plane with the reconstruction of the Cherenkov ring performed using the MAPMT information only.

To do this, the Cherenkov rings are reconstructed by minimizing the quantity 
\begin{equation}
S(R, X_C, Y_C) = \sum_{i=1}^{N_{pe}} [ (x_i - X_C)^2 + (y_i - Y_C)^2 - R^2 ]^2
\label{eq:fit3par}
\end{equation}
where $(x_i, y_i)$ are the coordinates of the $i_{th}$ photon hit, $(X_C, Y_C)$ are the coordinate of the ring center and $R$ its radius.
The event-by-event minimization of eq. (\ref{eq:fit3par}) with MINUIT \cite{minuit} can be performed either to determine the three free parameters $X_C$, $Y_C$ and $R$ ({\it 3par} fit) or to fit the radius only ({\it 1par} fit) fixing $X_C$ and $Y_C$ by using the GEM information.

The four GEM alignment constants have been determined by minimizing the quantity:
\begin{equation}
Q(X_{0}, Y_{0}, X_{1}, Y_{1}) = \sigma^2_R + \bar{(\Delta X)^2} + \bar{(\Delta Y)^2}
\end{equation}
where $\sigma^2_R$ is the RMS of the distribution of the Cherenkov ring radius in the {\it 1par} fit and $\bar{\Delta X}$ and $\bar{\Delta Y}$ are the mean values of the distributions of the difference between the GEM track (projected on the MAPMT plane) and the ring center coordinates obtained in the {\it 3par} fit.

\begin{figure}
\begin{center}
\resizebox{0.35\textwidth}{!}{%
  \includegraphics{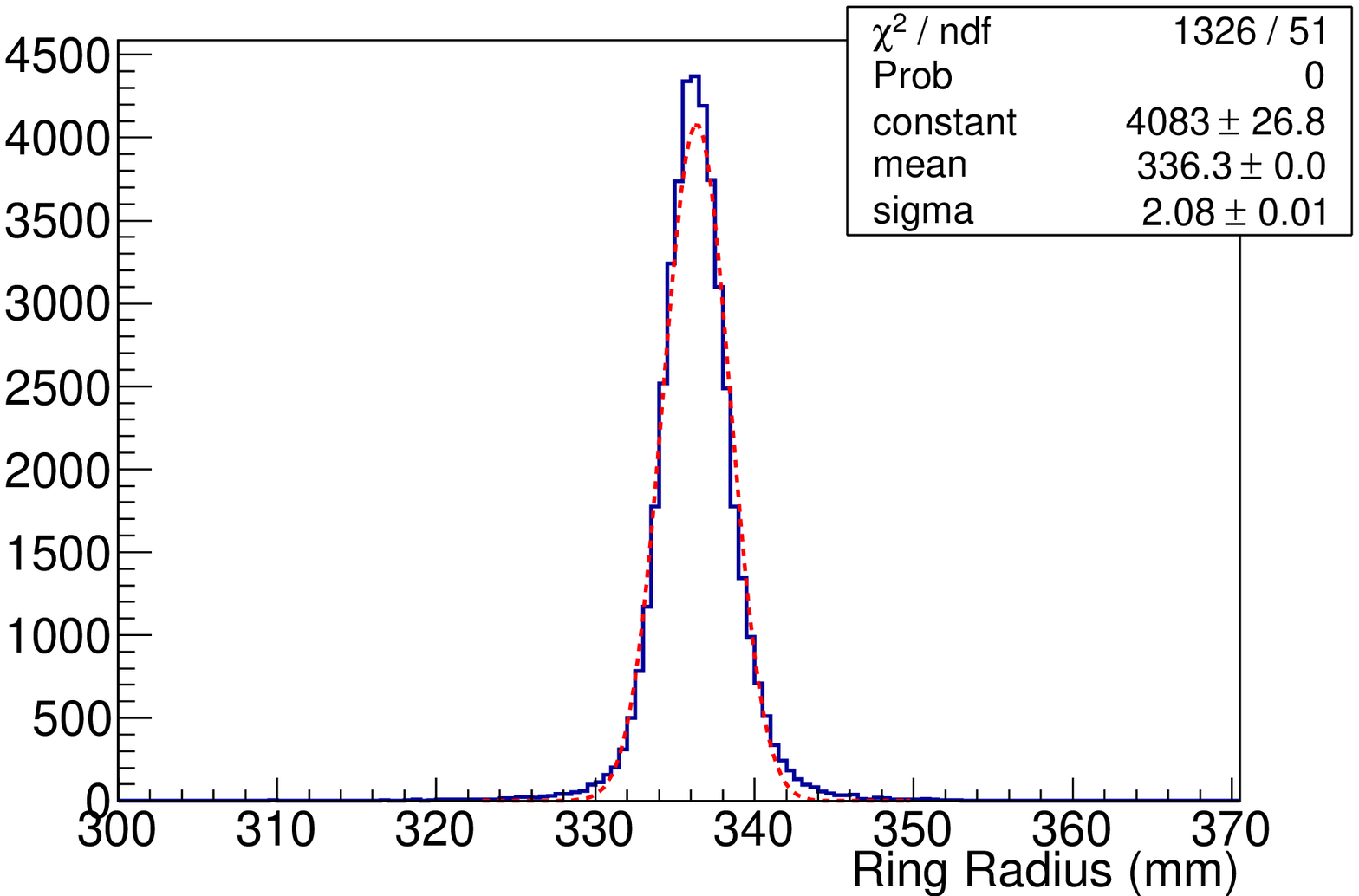}
}
\resizebox{0.35\textwidth}{!}{%
  \includegraphics{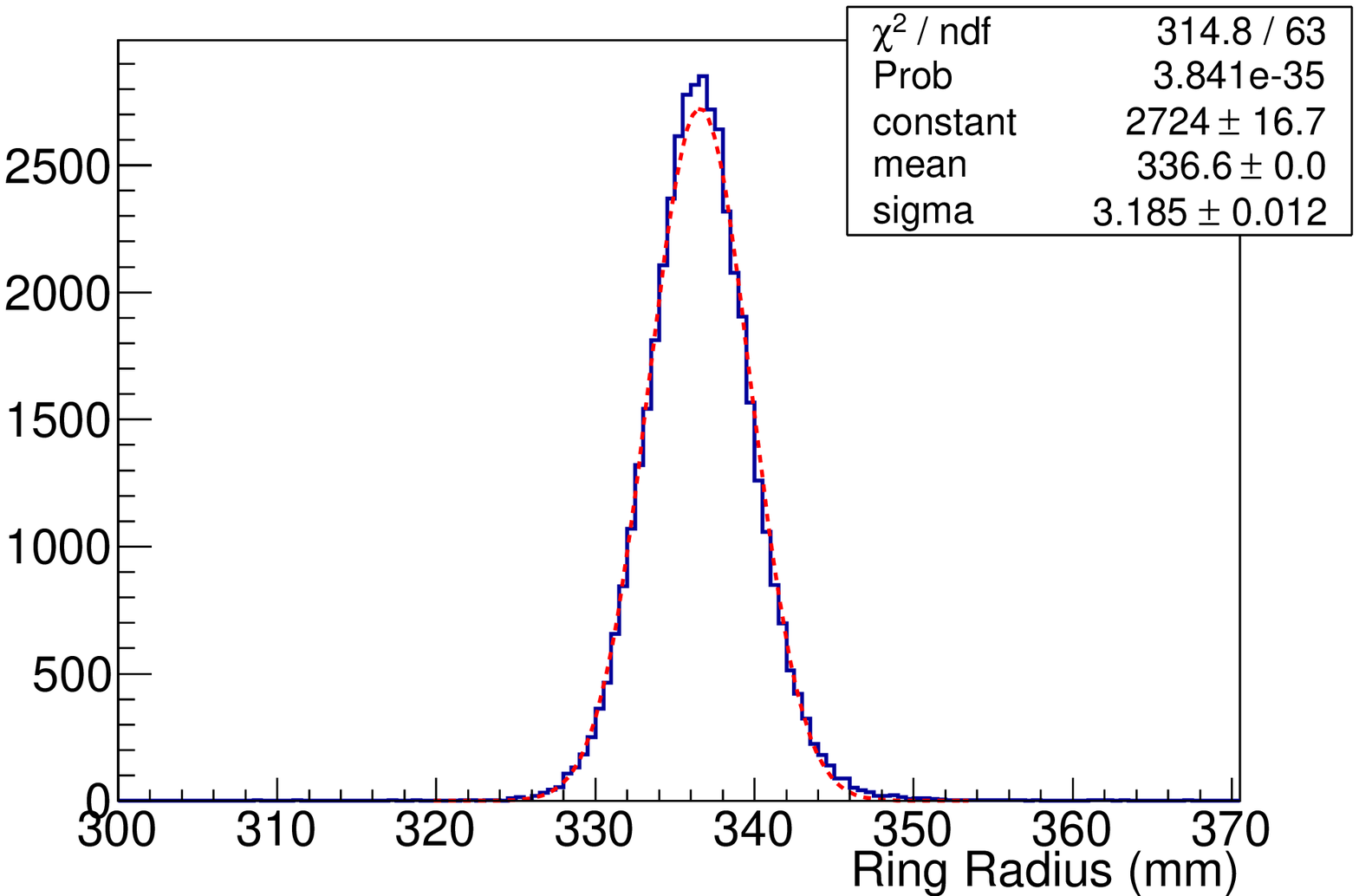}
}
\resizebox{0.35\textwidth}{!}{%
  \includegraphics{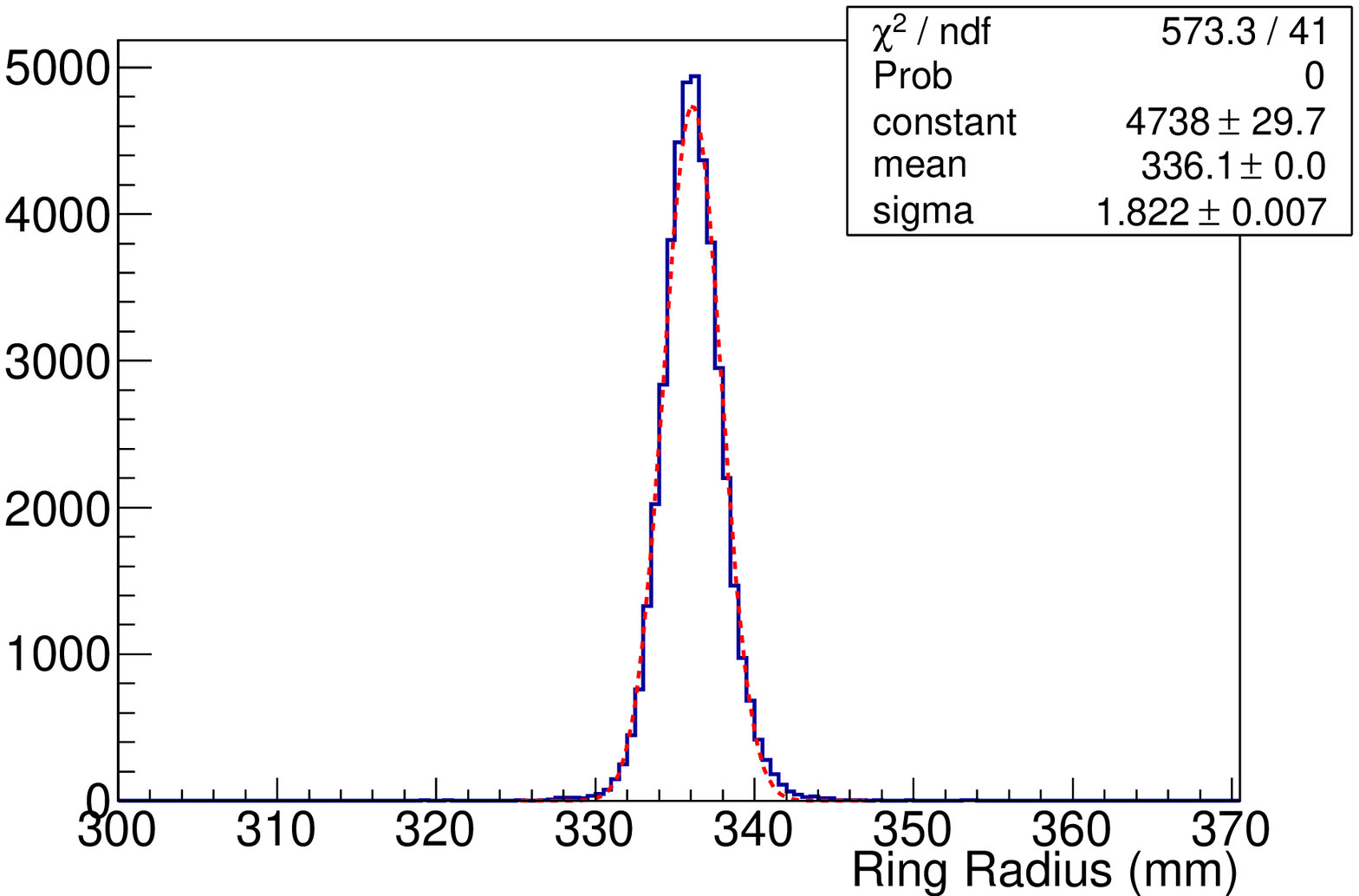}
}

\caption{Cherenkov ring radius from the {\it 3par} fit (top), from the {\it 1par} fit before (center) and after (bottom) the GEM alignment procedure. Data for 8 GeV/c pions on aerogel with $n=1.05$ and $t=2$ cm.}
\label{fig:AlignFit}
\end{center}
\end{figure}

In Fig. \ref{fig:AlignFit}, we show an example of the Cherenkov ring radius distribution from the {\it 3par} fit (top, $\sigma_R = 2.1$ mm), from the {\it 1par} fit before (center, $\sigma_R = 3.2$ mm) and after (bottom, $\sigma_R = 1.8$ mm) the offline alignment.
An improvement of more than $10 \%$ in the resolution has been obtained from the initial {\it 3par} to the final {\it 1par} fit at the end of the alignment procedure.
The stability of the procedure has been checked by calculating the alignment constants from several runs with different beam energies across the data taking period.
No run dependence of these constants has been found.

\begin{figure}
\begin{center}
\resizebox{0.45\textwidth}{!}{%
  \includegraphics{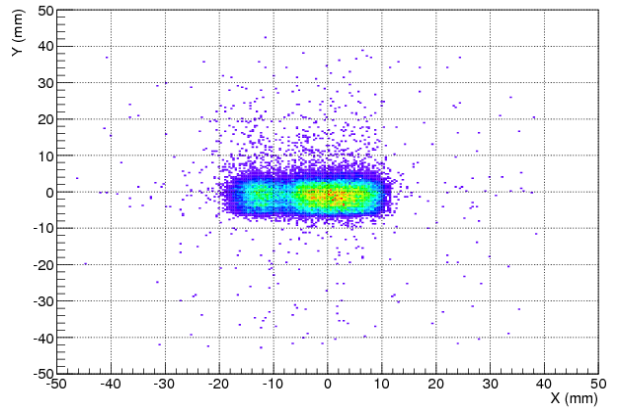}
}

\caption{Beam spot measured with the two GEMs and projected onto the RICH photon detector plane. Data for 8 GeV/c beam.}
\label{fig:BeamSpot}
\end{center}
\end{figure}

After the alignment, the beam spot size projected onto the MAPMT plane is shown in Fig. \ref{fig:BeamSpot}.
The different size between the horizontal and the vertical dimensions (15 mm vertically and 25 mm horizontally) is due to the different collimation.
The small gap at $x \approx -8$ mm comes from the noisy strips removed from the downstream GEM.

\subsection{MAPMT position correction}

\begin{figure}
\begin{center}
\resizebox{0.45\textwidth}{!}{%
  \includegraphics{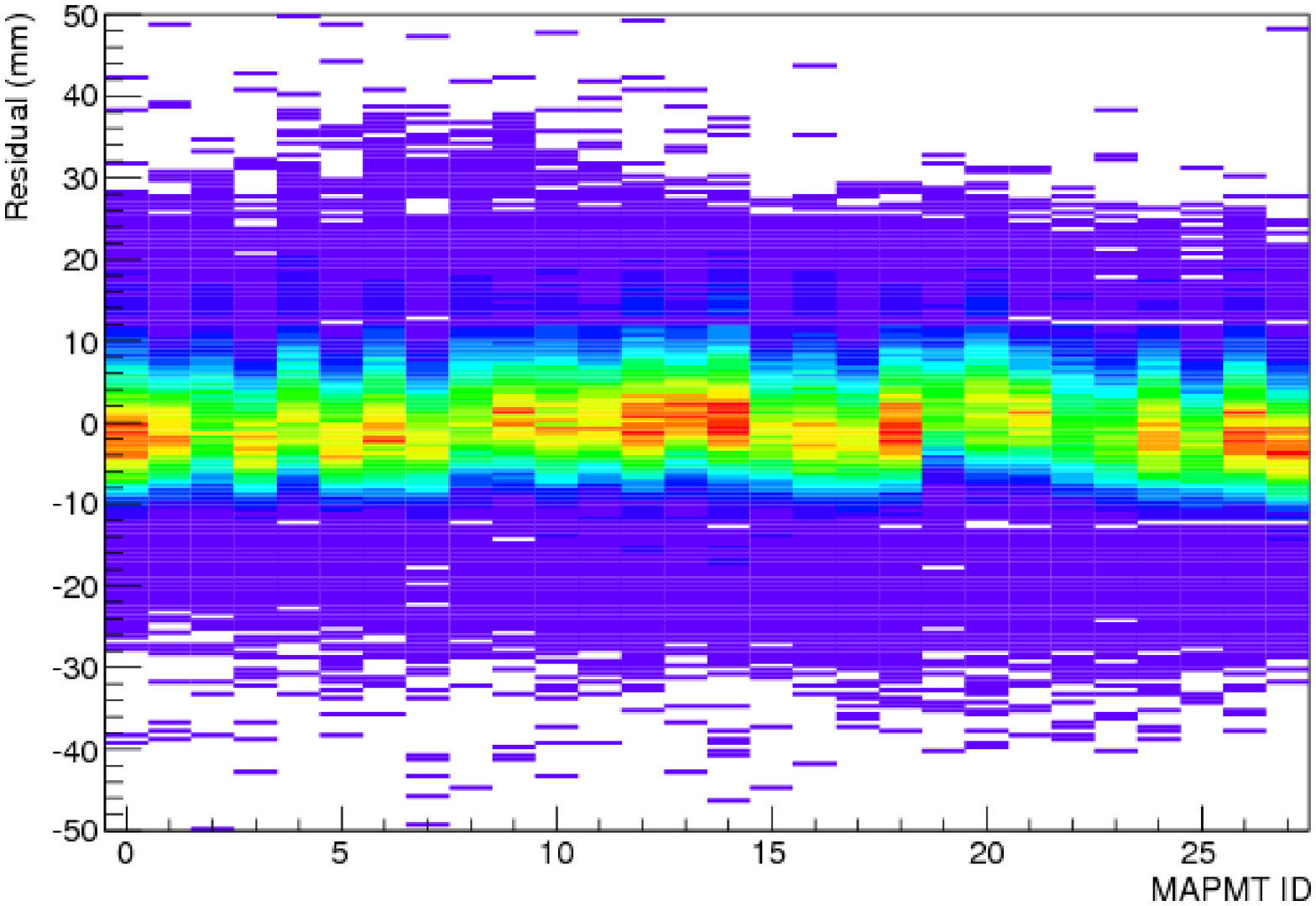}
}
\resizebox{0.45\textwidth}{!}{%
  \includegraphics{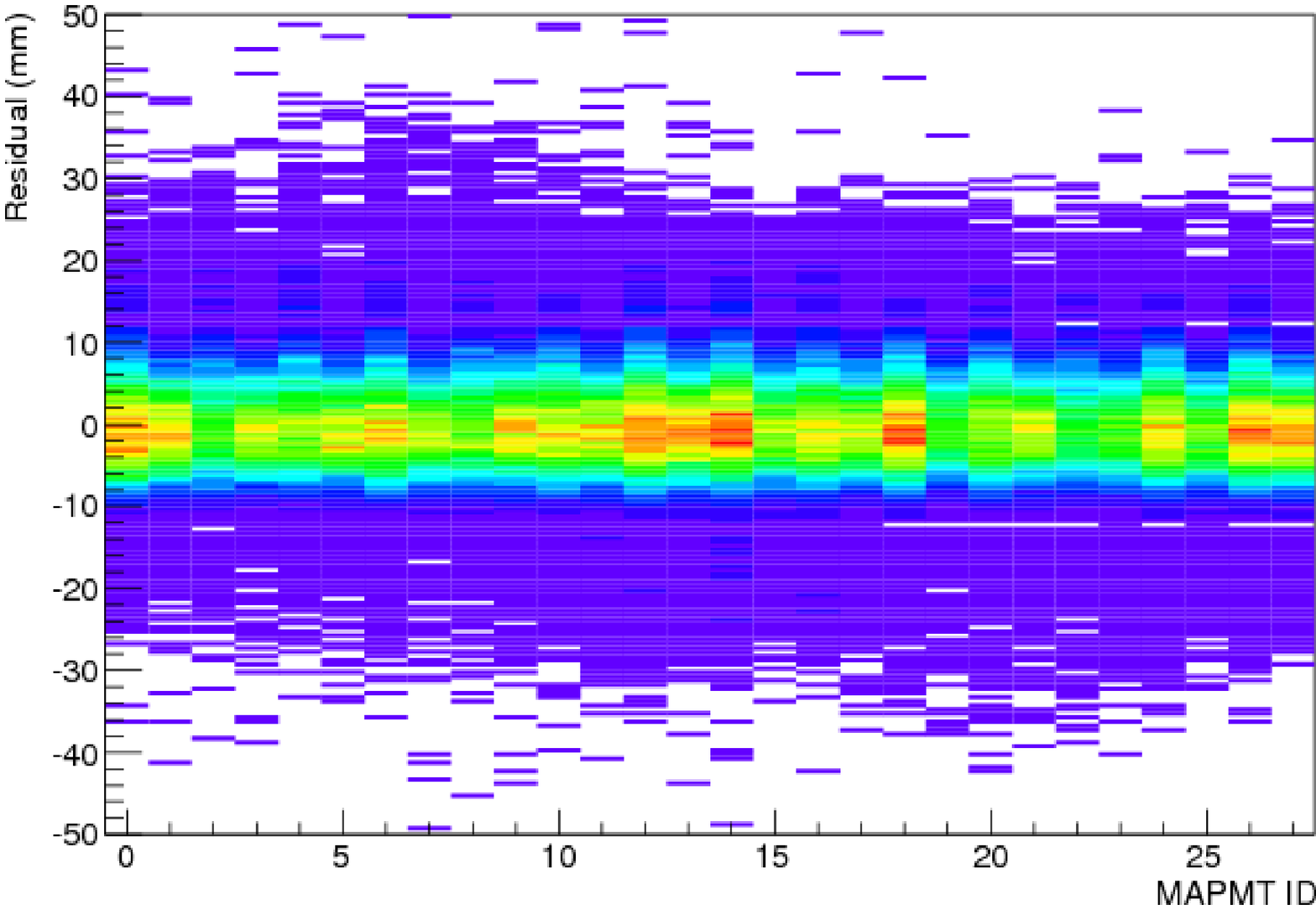}
}

\caption{Hit residual distributions for the 28 MAPMTs before (top plot) and after (bottom plot) the position correction.}
\label{fig:ResidualVsPmt}
\end{center}
\end{figure}

For each aerogel refractive index, the position of the MAPMTs on the support was moved according to the expected Cherenkov ring radius and was periodically measured during the data taking.
Finer alignment of the MAPMTs position was performed off-line by looking at the residuals of the photon hits, i.e. the distance of each hit from the fitted Cherenkov ring.
In the top plot of Fig. \ref{fig:ResidualVsPmt}, we show an example of the residual distributions for the 28 MAPMTs.
For each MAPMT, the shift from zero of the mean residual represents the correction to be applied to the measured position.
Typically, this correction amounts to a few mm and never exceeds 6 mm, the size of the MAPMT pixel.
The distributions of the residuals after the alignment procedure are shown in the bottom plot of Fig. \ref{fig:ResidualVsPmt}.
The different widths of the residual distributions between the even (H8500C-03) and odd (H8500C) MAPMTS is due to the UV photons, as will be discussed in Section \ref{sec:UVData}.

The alignment corrections have been computed for each group of runs taken in the same experimental conditions.
On average, the MAPMT position correction provides an improvement of the Cherenkov angle resolution of a few percent.

\subsection{Cherenkov event reconstruction}

\begin{figure}
\begin{center}
\resizebox{0.5\textwidth}{!}{%
  \includegraphics{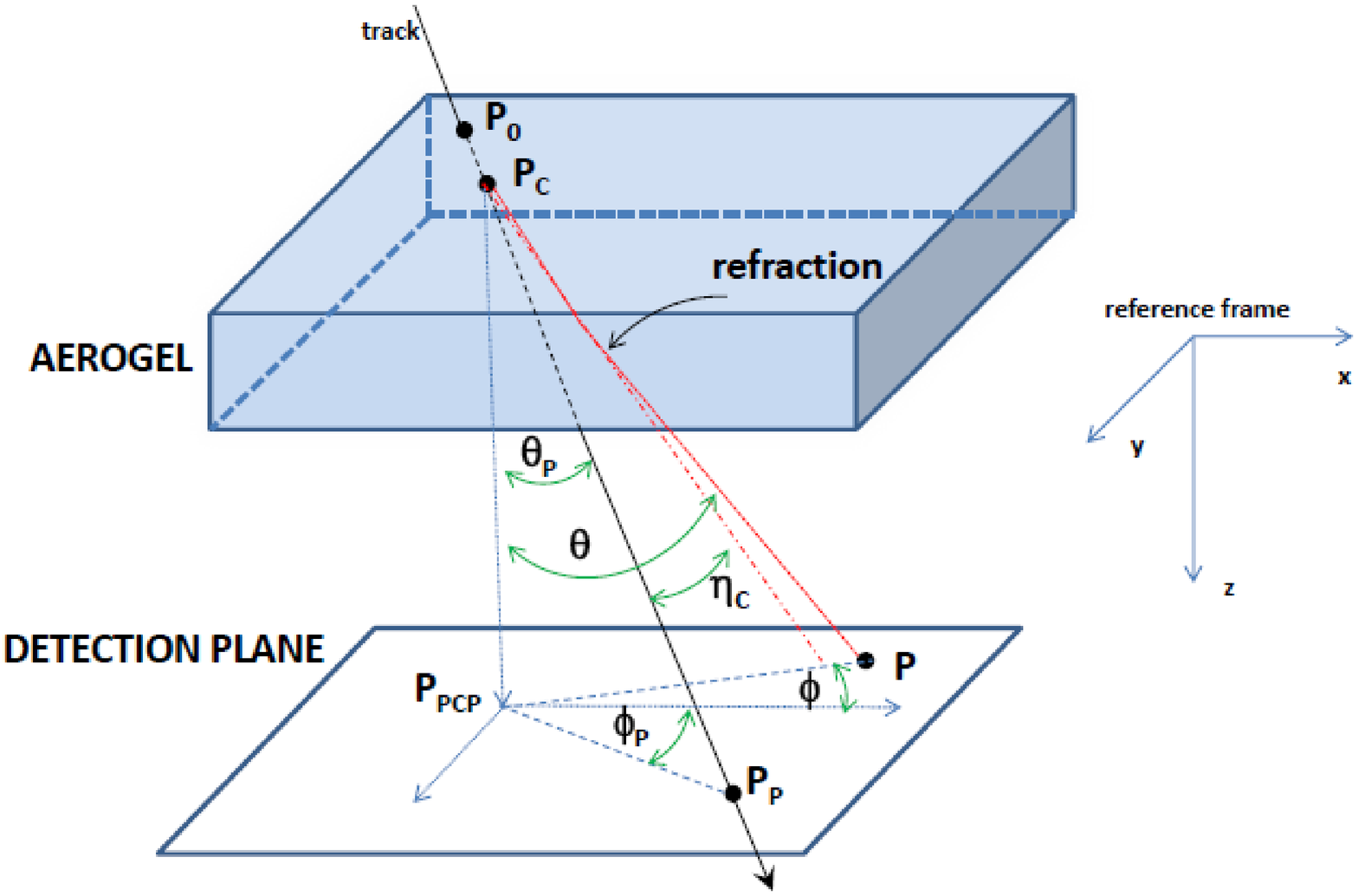}
}
\caption{Illustration of the Cherenkov angle reconstruction.}
\label{fig:AngleRec}
\end{center}
\end{figure}

The reconstruction of the Cherenkov angle has been performed following the method developed for the proximity RICH detectors of the ALICE HMPID \cite{ALICE} and the JLab Hall A \cite{JLab_HallA} RICHs. 
The method, illustrated in Fig. \ref{fig:AngleRec}, exploits the information of the tracking system and of the MAPMT array and provides, for a given photon hit, the Cherenkov angle, correctly taking into account the small slope of the hadron track, of the order of few mrad, which produces a non circular hit distribution on the MAPMT plane.

The reference frame has the $z$ axis along the beam line, the $y$ axis is vertical and pointing upwards and then $\hat{x} = \hat{y} \times \hat{z}$.
The thickness of the radiator is $t_{rad}$, while $t_{gap} = 994$ mm and $n_{air} = 1.000273$ are the length and refractive index of the air gap between the radiator and the photodetectors.
The track enters the radiator at the point $P_0$ and its projection onto the MAPMT plane is $P_p$.
Its polar and azimuthal angles are $\theta_p$ and $\phi_p$, respectively.

The Cherenkov photon is emitted at the point $P_c$, at a depth $L$ inside the radiator, and is detected at the point $P$ after the refraction when exiting from the aerogel.
In principle, a recursive calculation would allow a rather precise determination of $L$.
However, given the small thickness of the radiators with respect to the total gap length, we assumed $L = t_{rad} / 2$.

The point $P_{pcp}$ is the projection of the emission point onto the MAPMT plane.
The polar and azimuthal angles of the photon hit are $\theta$ and $\phi$, while the Cherenkov angle, i.e. the angle between the photon direction and the track, is $\eta_c$.
It is important to note that the refraction in the passage from the aerogel to the air changes its polar angle but not the azimuthal angle.
As a result, the photon is moving on a plane and, by using the Snell equation, if the angles are known one can calculate the ring radius $R$ as

\begin{equation}
R = (t_{rad} -L) \tan(\theta) + t_{gap} \frac{n \sin(\theta)}{\sqrt{n_{air}^2 - n^2 \sin^2(\theta)}} \equiv f(\theta)
\label{eq:AngleRec}
\end{equation}

\noindent
When the emission angle is not known but the radius $R$ is measured, this equation must be inverted to obtain $\theta$.
The problem is solved by minimizing the difference 
\begin{equation}
g(\theta) = (R - f(\theta))^2
\end{equation}
Once $\theta$ is obtained, the Cherenkov angle of the photon hit is given by
\begin{equation}
\cos(\eta_c) = \sin(\theta_p) \sin(\theta) \cos(\phi - \phi_p) + \cos(\theta_p) \cos(\theta)
\end{equation}
Finally, the Cherenkov angle of the event, where $m$ photon hits have been detected, is given by the average of the $m$ Cherenkov angles.

Knowing the Cherenkov angle, one can derive also the measured refractive index of the radiator
\begin{equation}
\label{eq:ref_ind}
n = \frac{1}{\beta \cos(\eta_c)}
\end{equation}
where the velocity of the particle $\beta$ is computed from the beam momentum and using the gas Cherenkov counter information to discriminate between pions and heavier hadrons.

An iterative algorithm of the Cherenkov ring reconstruction has been implemented in order to study background hits.
A hit is considered as background if its distance from the fitted ring is bigger than a given threshold value.
In this case, the hit is removed and a new fit is performed taking into account only the remaining good Cherenkov hits.
The threshold value has been optimized to the data: a too loose value will include too many background hits, thus worsening the resolution, while a too high value will remove good photons, again worsening the resolution.
An optimal value of 12 mm (i.e. twice the pixel size) has been found.

\section{Results}

\subsection{Cherenkov angle resolution studies with pions}
\label{sec:pions}

We discuss now the performances of the prototype in the nominal condition, i.e. pions with momentum of 8 GeV/c and an aerogel radiator with $n=1.05$ and $t_{rad} = 20$ mm thickness.
These are expected to be the most challenging conditions in the CLAS12 RICH. 
Pions are selected through the gas threshold Cherenkov signal, as shown in Fig. \ref{fig:CerAdc}.
We also require a minimum of four hits on the MAPMTs.

\begin{figure}
\begin{center}
\resizebox{0.5\textwidth}{!}{%
  \includegraphics{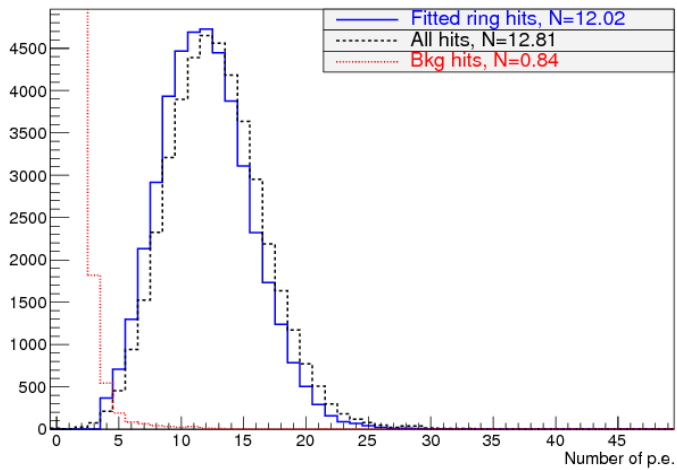}
}

\caption{ Total number of detected hits (dashed black line), the number of p.e. (full blue line) and the number of background hits (dotted red line) per event measured in the RICH prototype with 8 GeV/c momentum pions and aerogel with $n=1.05$ and $t_{rad}=20$ mm.}
\label{fig:Npe_pions}
\end{center}
\end{figure}

\begin{figure}
\begin{center}
\resizebox{0.5\textwidth}{!}{%
  \includegraphics{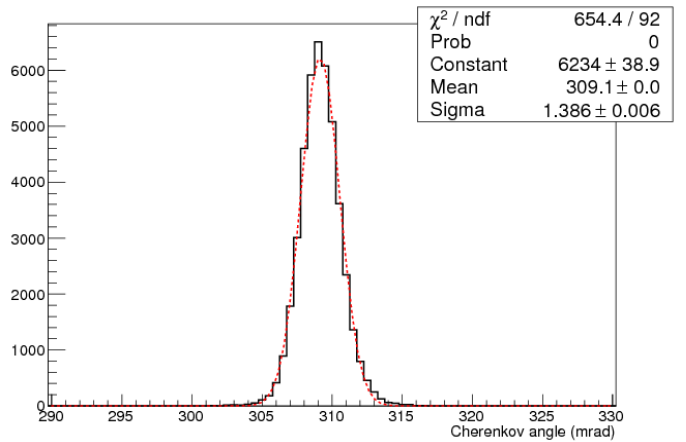}
}
\caption{Cherenkov angle distribution for 8 GeV/c momentum pions and aerogel with $n=1.05$ and $t_{rad}=20$ mm. The dashed line is a gaussian fit of the distribution, whose fitted parameters are shown in the table.}
\label{fig:Angle_pions}
\end{center}
\end{figure}

In Fig. \ref{fig:Npe_pions}, we show the total number of detected hits (dashed black line), the number $N_{pe}$ of photo-electrons (p.e.) for the reconstructed rings (full blue line) and the number $N_{bkg}$ of background hits (dotted red line).
We have $N_{pe} = 12.02 \pm 0.02$ p.e. per pion Cherenkov ring and $N_{bkg} \approx 1$.
The ring coverage of our RICH prototype was in this case about $70\%$, thus for a full coverage detection system we may estimate $N_{pe} \approx 17$.
The Cherenkov angle distribution is shown in Fig. \ref{fig:Angle_pions}.
The gaussian fit of the distribution yields $\eta_c = 309.14 \pm 0.01$ mrad and a resolution $\sigma_{\eta} = 1.39 \pm 0.01$ mrad.

\begin{figure}
\begin{center}
\resizebox{0.5\textwidth}{!}{%
  \includegraphics{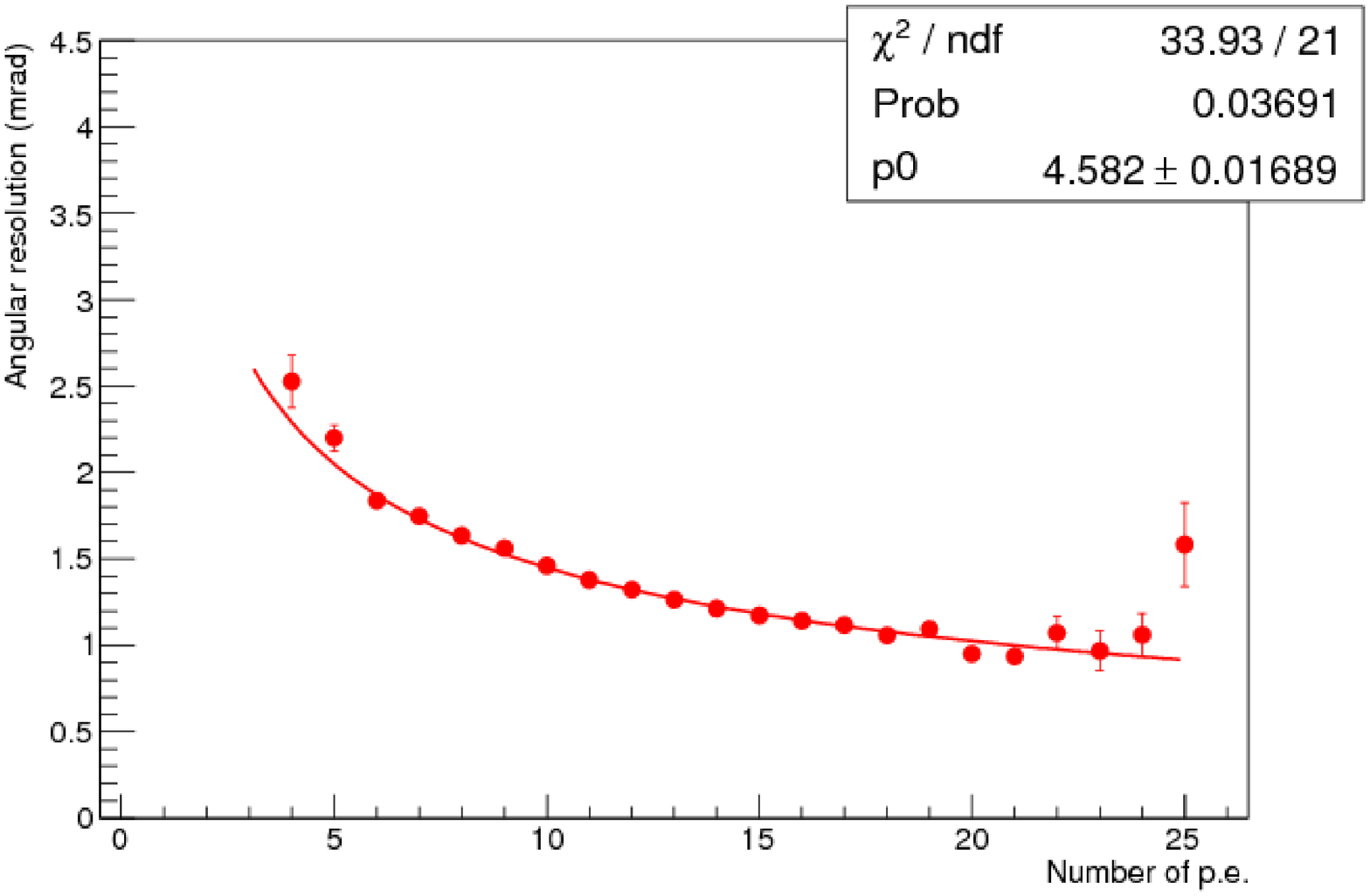}
}
\caption{Angular resolution as a function of the number of p.e. for 8 GeV/c momentum pions and aerogel with $n=1.05$ and $t_{rad}=20$ mm. The line is a fit with eq. (\ref{eq:SigmaFitFunction}), with $\sigma_{1pe} \equiv p_0$.}
\label{fig:ResolutionVsNpe_pions}
\end{center}
\end{figure}

As expected, no variations have been found in $\eta_c$ as a function of the number of photoelectrons: the slope of the $N_{pe}$ dependence is smaller than $10^{-2}$ mrad/$N_{pe}$.
On the other hand, the gaussian width of the Cherenkov angle distribution is expected to be proportional to the inverse of the squared root of $N_{pe}$:
\begin{equation}
\label{eq:SigmaFitFunction}
\sigma_{\eta} = \frac{\sigma_{1pe}}{\sqrt{N_{pe}}}
\end{equation}
where $\sigma_{1pe}$ is the single photon detection resolution.
The Cherenkov angle resolution as a function of the number of p.e. is shown in Fig. \ref{fig:ResolutionVsNpe_pions}.
We measured up to 25 p.e. per ring and the fit with eq. (\ref{eq:SigmaFitFunction}) gives the single photon resolution $\sigma_{1pe} = 4.58 \pm 0.02$ mrad.

As a test of the reconstruction algorithm, the $N_{pe}$ has been repeated by adding to eq. (\ref{eq:SigmaFitFunction}) a constant term.
The results showed a negligible constant term (below $0.1$ mrad) and basically no variation in the $\sigma_{1pe}$.
Thus, we can conclude that the possible experimental artifacts, as for example residual misalignment of the setup,  are well under control.

The pion reconstruction efficiency may be estimated by computing the ratio 
\begin{equation}
\epsilon_{\pi} = \frac{N_{\pi}^{RICH}}{N_{\pi}}
\end{equation}
where $N_{\pi}$ is the number of events with a gas Cherenkov counter signal above the threshold and $N_{\pi}^{RICH}$ represents the number of these events that are also reconstructed as pions in the RICH prototype.
An event in the RICH prototype is taken as a pion if the reconstructed Cherenkov angle falls in the range $\eta_c \pm 3 \sigma_{\eta}$.
We obtain $\epsilon_{\pi} = 98.8 \pm 0.1 \%$, with the small fraction of event loss mainly due to the events with less than 4 photoelectrons.

\subsection{The $\pi$ / K comparison}

\begin{figure}
\begin{center}
\resizebox{0.4\textwidth}{!}{%
  \includegraphics{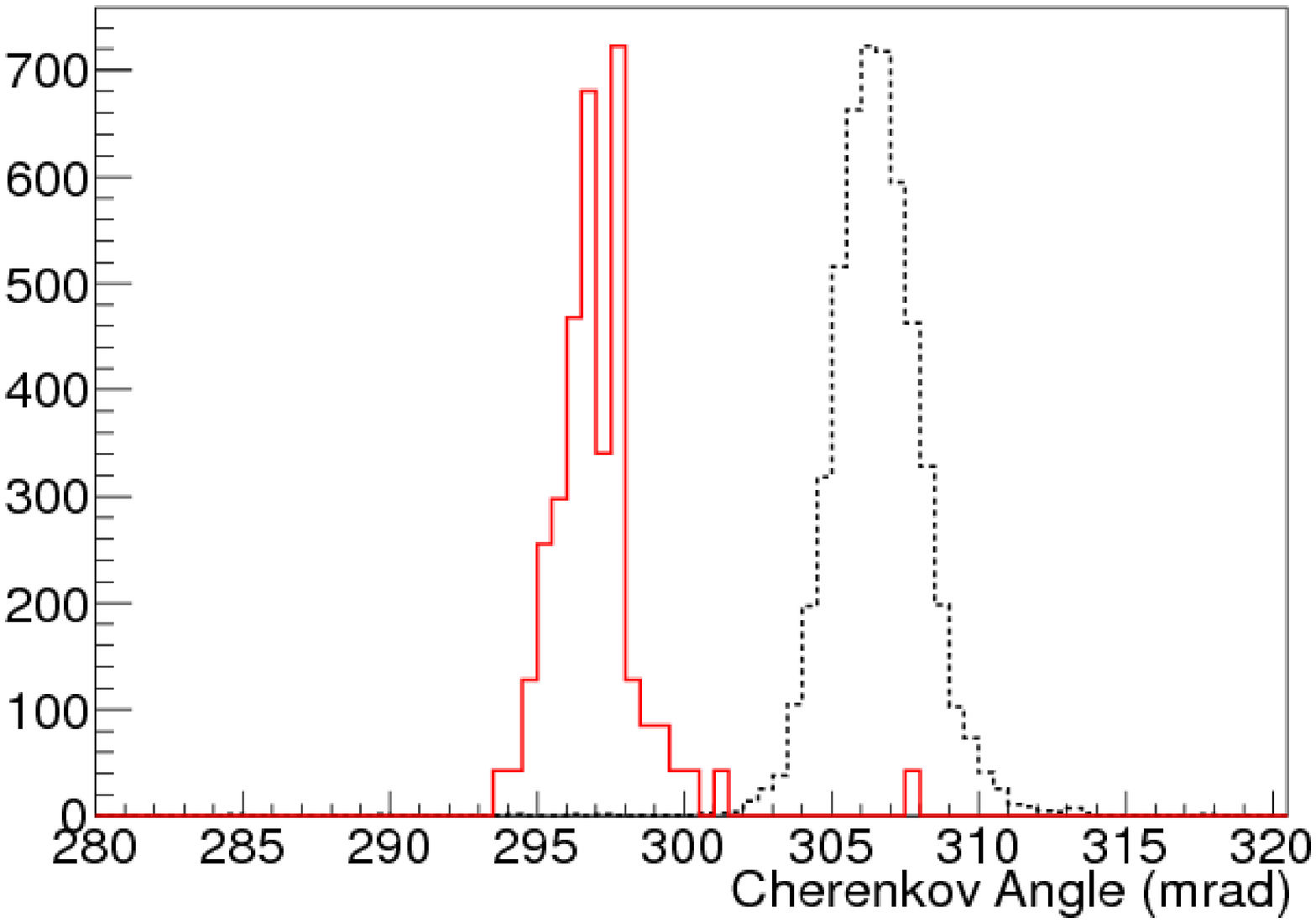}
}
\resizebox{0.4\textwidth}{!}{%
  \includegraphics{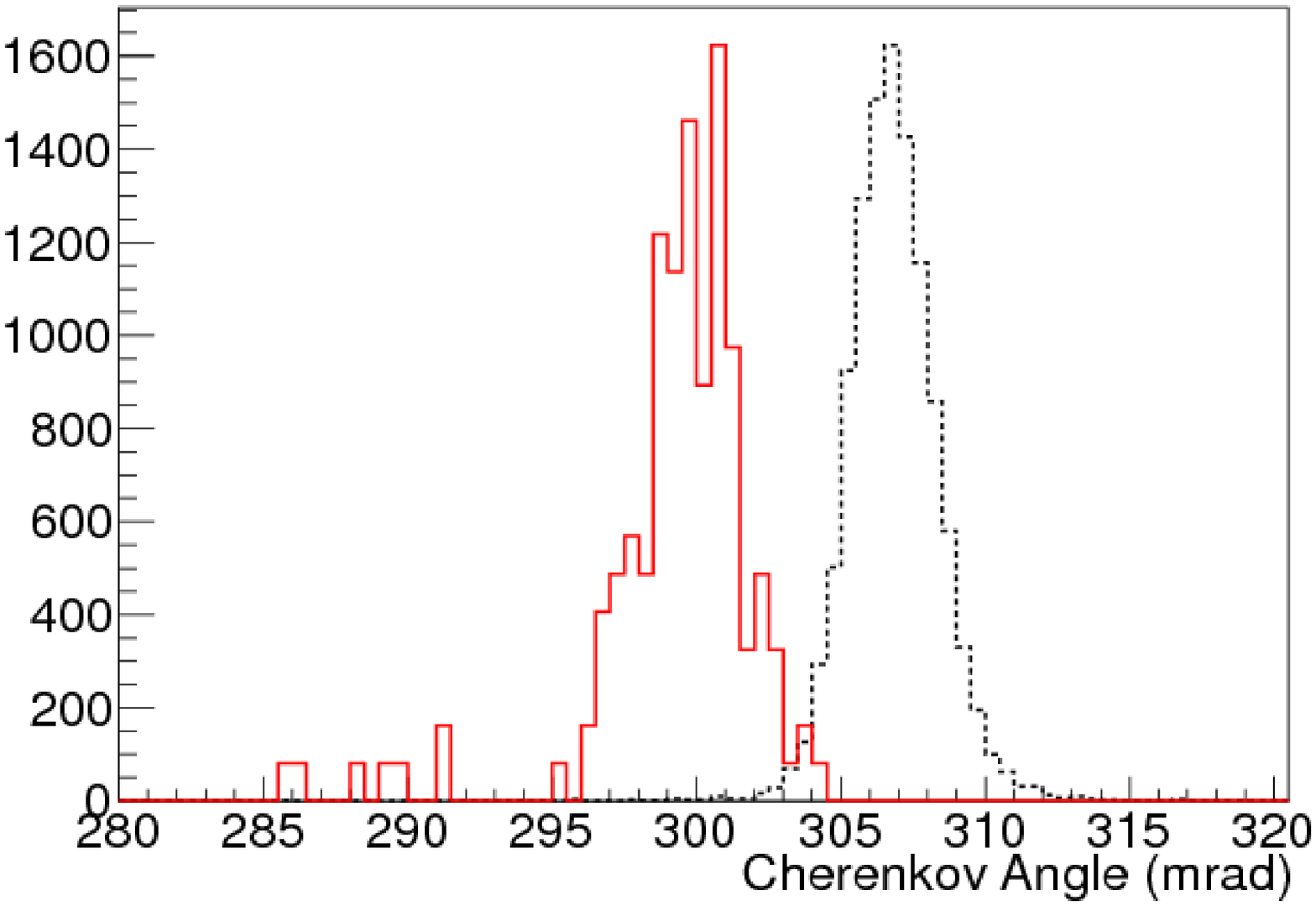}
}
\resizebox{0.4\textwidth}{!}{%
  \includegraphics{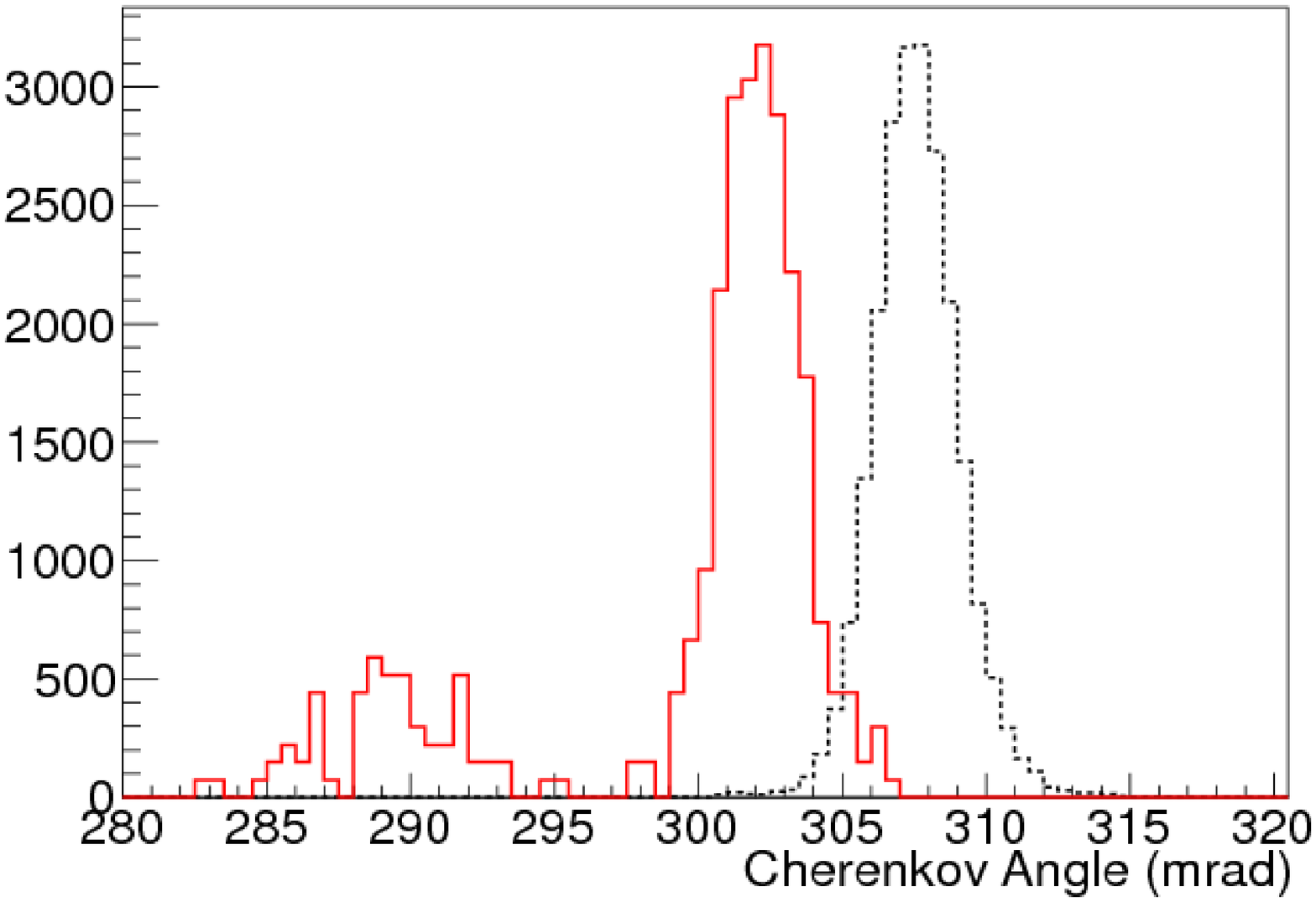}
}
\caption{Pion Cherenkov angle distributions (dashed black histograms) compared with those from events with gas Cherenkov signal below threshold (full red histograms, arbitrarily rescaled), for $P=6,7,8$ GeV/c beam (from top to bottom). Data for aerogel with $n=1.05$ and $t_{rad}=20$ mm.}
\label{fig:Angles}
\end{center}
\end{figure}

Events below the gas Cherenkov counter threshold include both kaons and antiprotons.
In Fig. \ref{fig:Angles}, we compare the Cherenkov angle distributions measured by the RICH prototype for these events (full red histograms) with the distributions for pions (dashed black histograms) at the three beam momenta $P=6,7,8$ GeV/c, from top to bottom.
The full red histograms are arbitrarily rescaled to make them visible.
We see the prominent kaon peak separated from the pion one and, at the highest energy, some tail of the $\bar{p}$ peak, which disappears as the energy decreases because it goes out of the MAPMTs radial coverage.

\begin{figure}
\begin{center}
\resizebox{0.35\textwidth}{!}{%
  \includegraphics{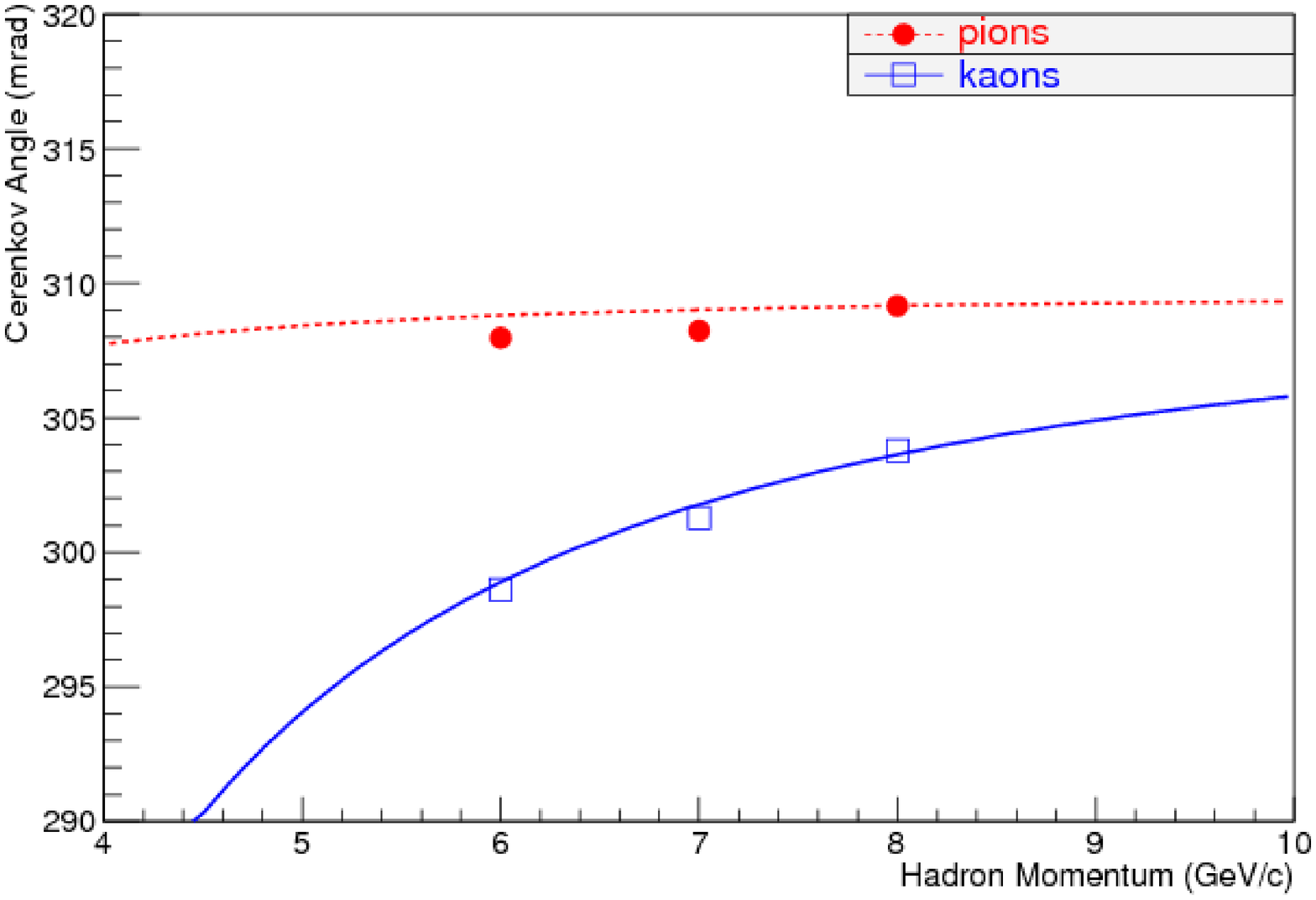}
}
\resizebox{0.35\textwidth}{!}{%
  \includegraphics{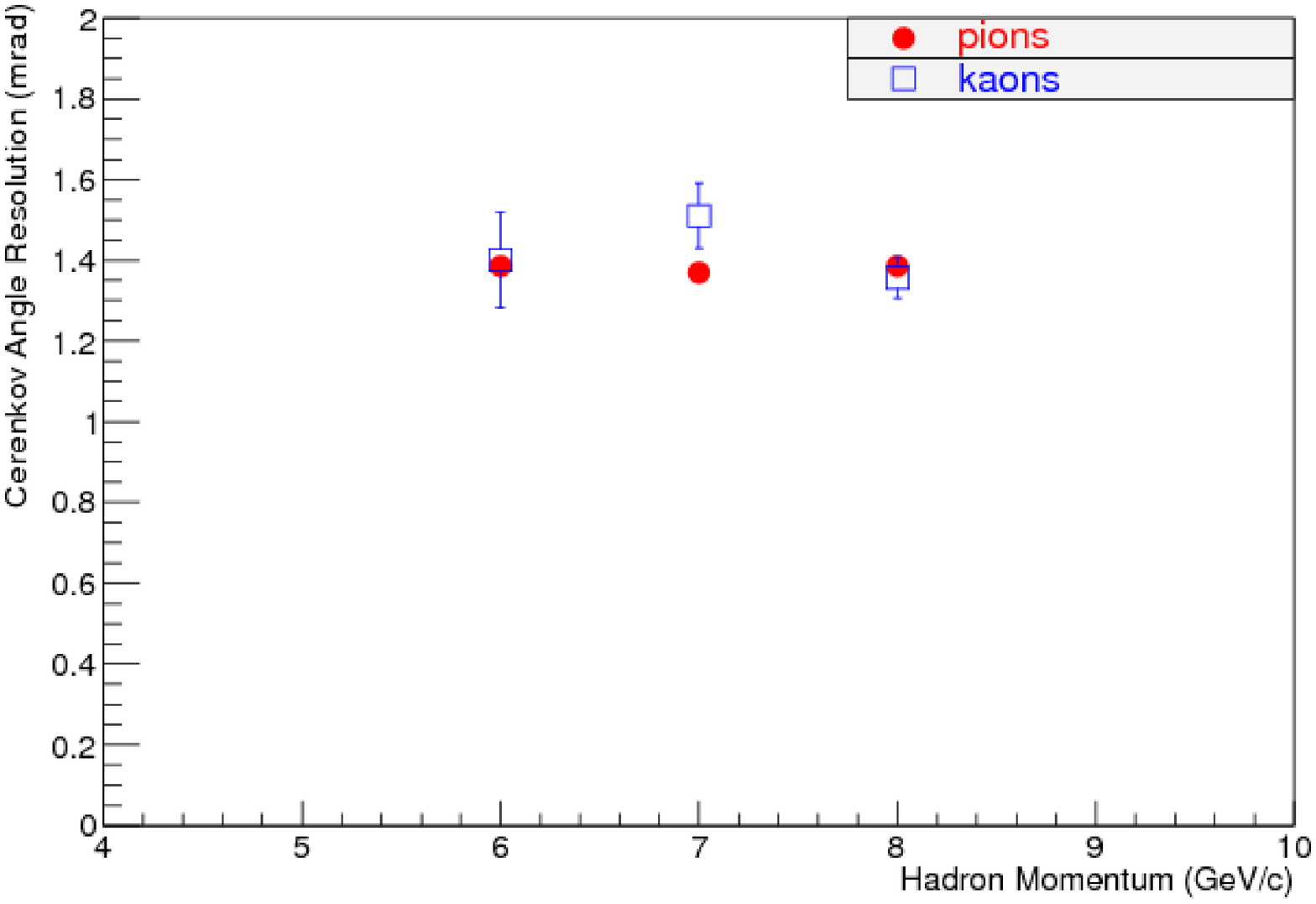}
}
\caption{Top plot: mean value of the Cherenkov angle measured for pions (full red circles) and kaons (empty blue squares) as a function of the hadron momentum, compared with the corresponding expected dependencies (dashed and full lines, respectively). Bottom plot: the same for the angular resolution. Data for aerogel with $n=1.05$ and $t_{rad}=20$ mm.}
\label{fig:AngleVsMom}
\end{center}
\end{figure}

Gaussian fits of these distributions provide the Cherenkov angle and resolution for the different hadrons.
In Fig. \ref{fig:AngleVsMom}, we show the gaussian mean (top plot) and sigma (bottom plot) of the Cherenkov angle distributions as a function of the hadron momentum, for pions (full red circles) and kaons (empty blue squares).
The measured mean values agree well with the calculated dependencies of eq. (\ref{eq:ref_ind}), shown by the dashed and full lines, respectively.
No momentum dependence has been found for the pion resolutions.
For the kaons, we observed some fluctuations due to the smaller statistics of the data sample, however the fitted values are the same as for the pions within about $1 \sigma$.

The number of sigma separation between the pion and the kaon Cherenkov angle peaks can be computed using the standerd gaussian estimator
\begin{equation}
\label{eq:separation}
n_{\sigma} = \frac{\eta_C^{\pi} - \eta_C^{K}}{ [\sigma_{\eta}^{\pi} + \sigma_{\eta}^{K}] / 2}.
\end{equation}

\noindent
We obtain $n_{\sigma} \ge 4$ in the whole momentum range, corresponding to a pion rejection factor of about 1:500, as required for the CLAS12 RICH.
Non-gaussian tails in the Cherenkov angle distributions could reduce the real rejection power.
Part of the tails are due to non optimal tracking reconstruction (in particular events with more than one track).
In addition, events were accepted starting from a minimum number of photoelectrons equal to 4. 
This very low number could be increased up to more realistic values of 6 or 7 with a minor expense of less than 10\% of efficiency. 
The requirement of one track and at least 7 p.e. provides a 1:500 rejection power as measured on the real distribution of tagged pions.

\subsection{UV photon resolution}
\label{sec:UVData}

\begin{figure}
\begin{center}
\resizebox{0.35\textwidth}{!}{%
  \includegraphics{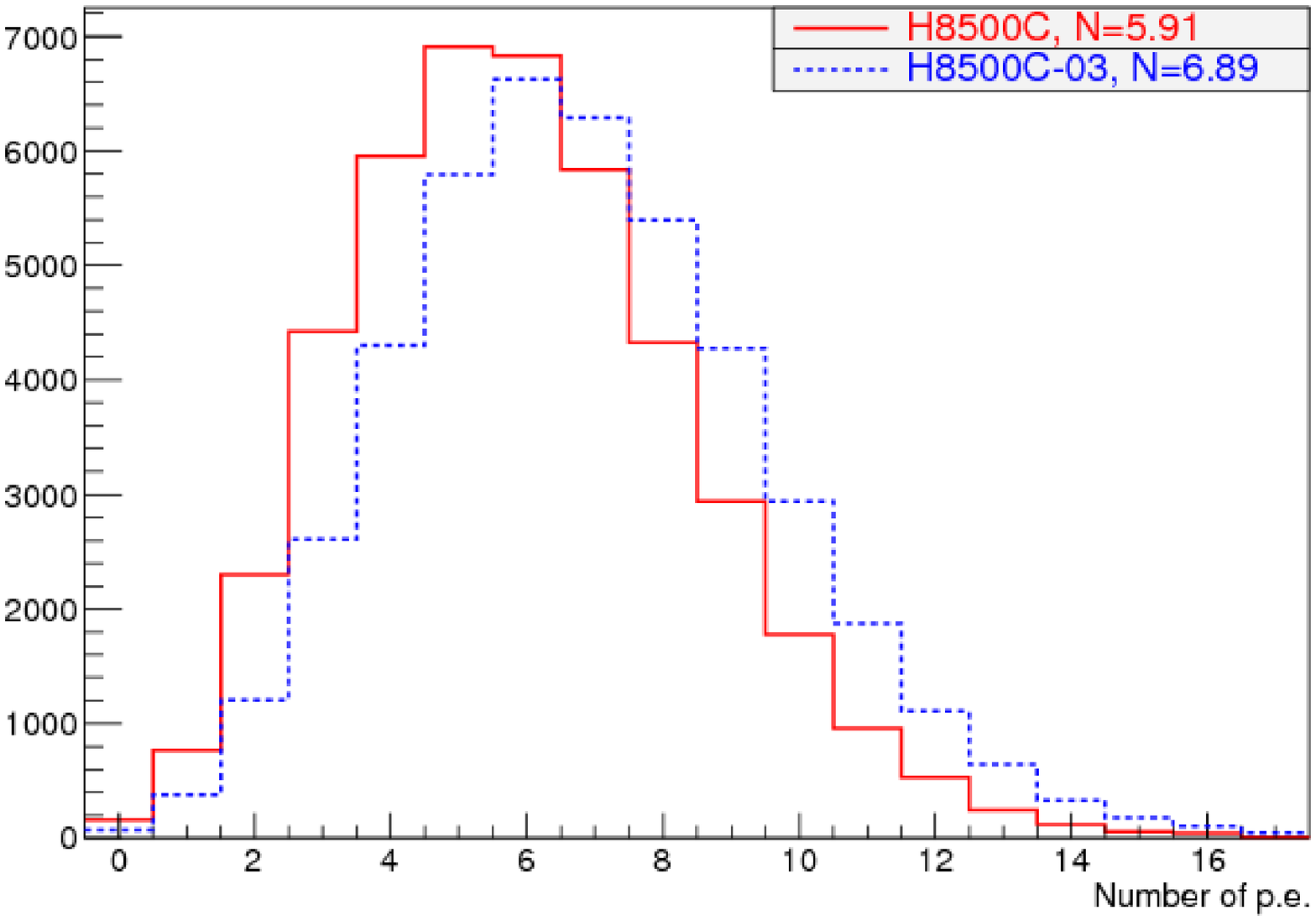}
}
\resizebox{0.35\textwidth}{!}{%
  \includegraphics{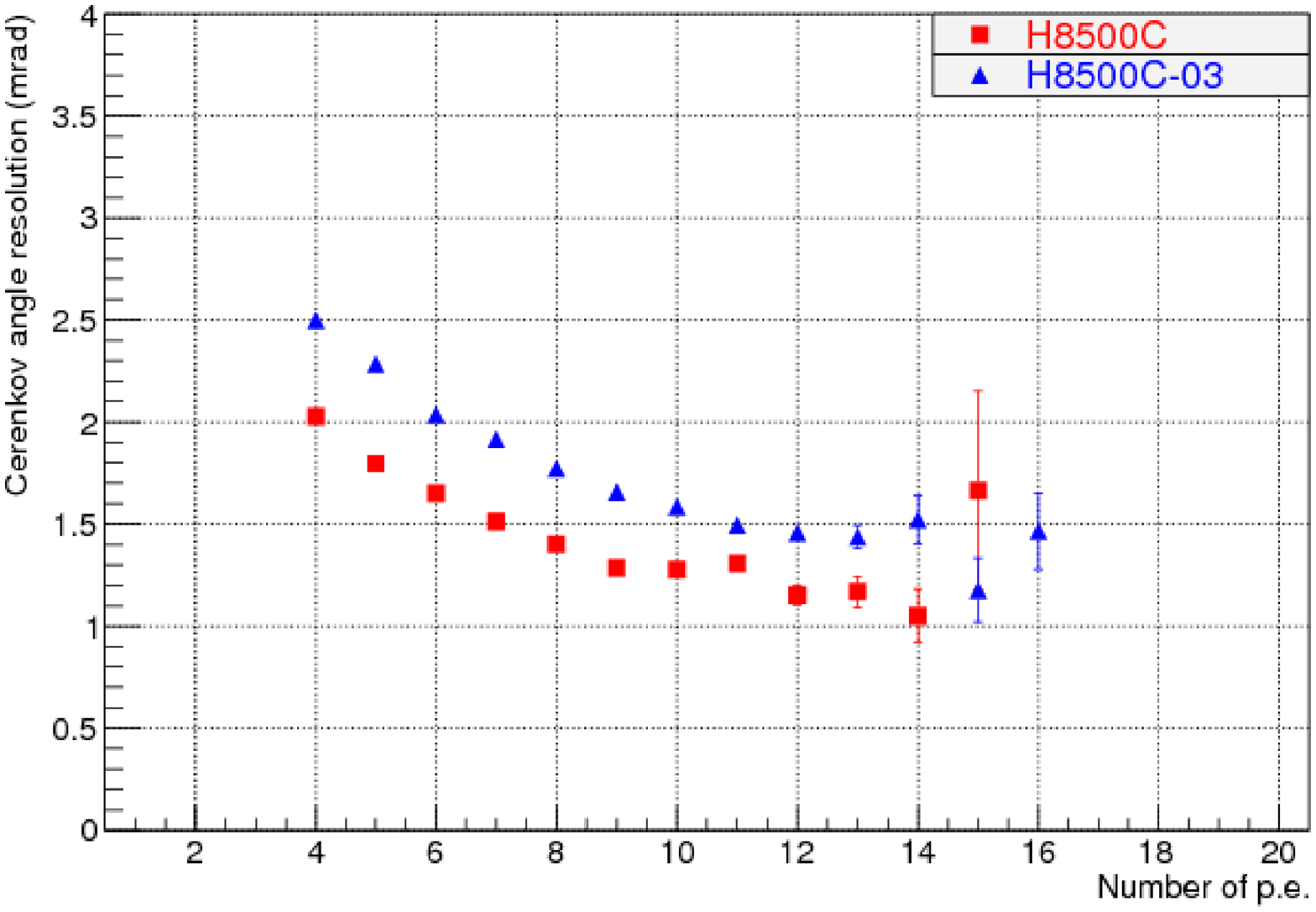}
}
\caption{Top plot: Number of p.e. per Cherenkov ring measured with the fourteen H8500C (red full line) and with the fourteen H8500C-03 (blue dashed line).
Bottom plot: Cherenkov angle resolution as a function of $N_{pe}$ measured with the fourteen H8500C (red squares) and with the fourteen H8500C-03 (blue triangles) MAPMTs.
Data for 8 GeV/c momentum pions and aerogel with $n=1.05$ and $t_{rad}=20$ mm.}
\label{fig:UV}
\end{center}
\end{figure}

The array of MAPMTs includes 14 normal glass window H8500C and 14 UV-enhanced glass window H8500C-03, alternated along the ring.
The effect of the UV photons can be estimated by comparing the photon yield and Cherenkov angle resolution using only the MAPMTs of the same type.
In Fig. \ref{fig:UV}, we report this comparison for pions with a momentum of 8 GeV/c.
The top plot shows the number of p.e. per event.
We see that the H8500C-03 MAPMTs detect about one hit per event more than the H8500C MAPMTs.
The bottom plot shows the Cherenkov angle resolution as a function of $N_{pe}$.
In spite of seeing more light, the H8500C-03 provide a worse resolution, because UV photons suffer more scattering inside the aerogel and in the air gap and also because of the larger chromatic dispersion (see Sect. \ref{sec:ChrData}).
Fitting the data with eq. (\ref{eq:SigmaFitFunction}), we extracted $\sigma_{1pe} = 4.01 \pm 0.01$ mrad for the H8500C and  $\sigma_{1pe} = 5.03 \pm 0.02$ mrad for the H8500C-03.

Given these results, the H8500C MAPMTs have been chosen for the CLAS12 RICH application.

\subsection{Aerogel refractive index}

\begin{figure}
\begin{center}
\resizebox{0.35\textwidth}{!}{%
  \includegraphics{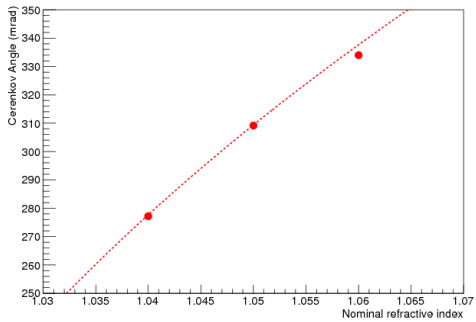}
}
\caption{Cherenkov angle measured with aerogel tiles of of thickness $t=20$ mm and different nominal refractive index (full circles) compared with the expected values (dashed line).
Data for pions with 8 GeV/c momentum.}
\label{fig:AngleVsRefInd}
\end{center}
\end{figure}

Measurements have been performed with aerogel tiles with nominal refractive index $n=1.04, 1.05, 1.06$ and thickness $t_{rad}=20$ mm.
In Fig. \ref{fig:AngleVsRefInd}, we show the Cherenkov angle measured for the three refractive indices (full circles) compared with the expected dependence (curve).
Deviations from the expected value, represented by the curve, indicate a difference of the actual refractive index from the nominal one.
For the two smaller refractive indices, the measured difference is below $0.1 \%$, while for $n=1.06$ a difference of the order of $0.2 \%$was found, within (although close to) the typical tolerance achievable with the present production technology. 

\begin{figure}
\begin{center}
\resizebox{0.35\textwidth}{!}{%
  \includegraphics{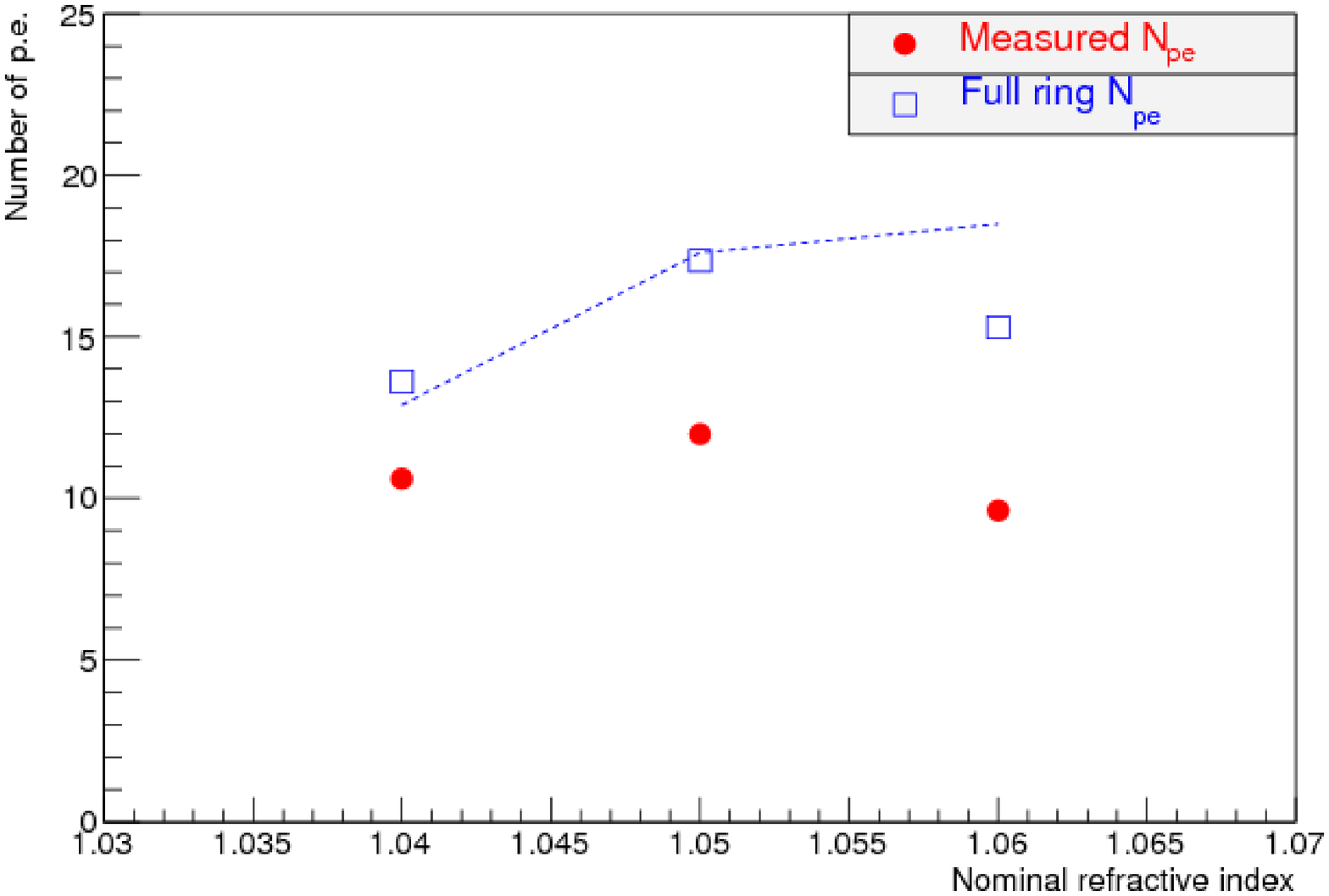}
}
\resizebox{0.35\textwidth}{!}{%
  \includegraphics{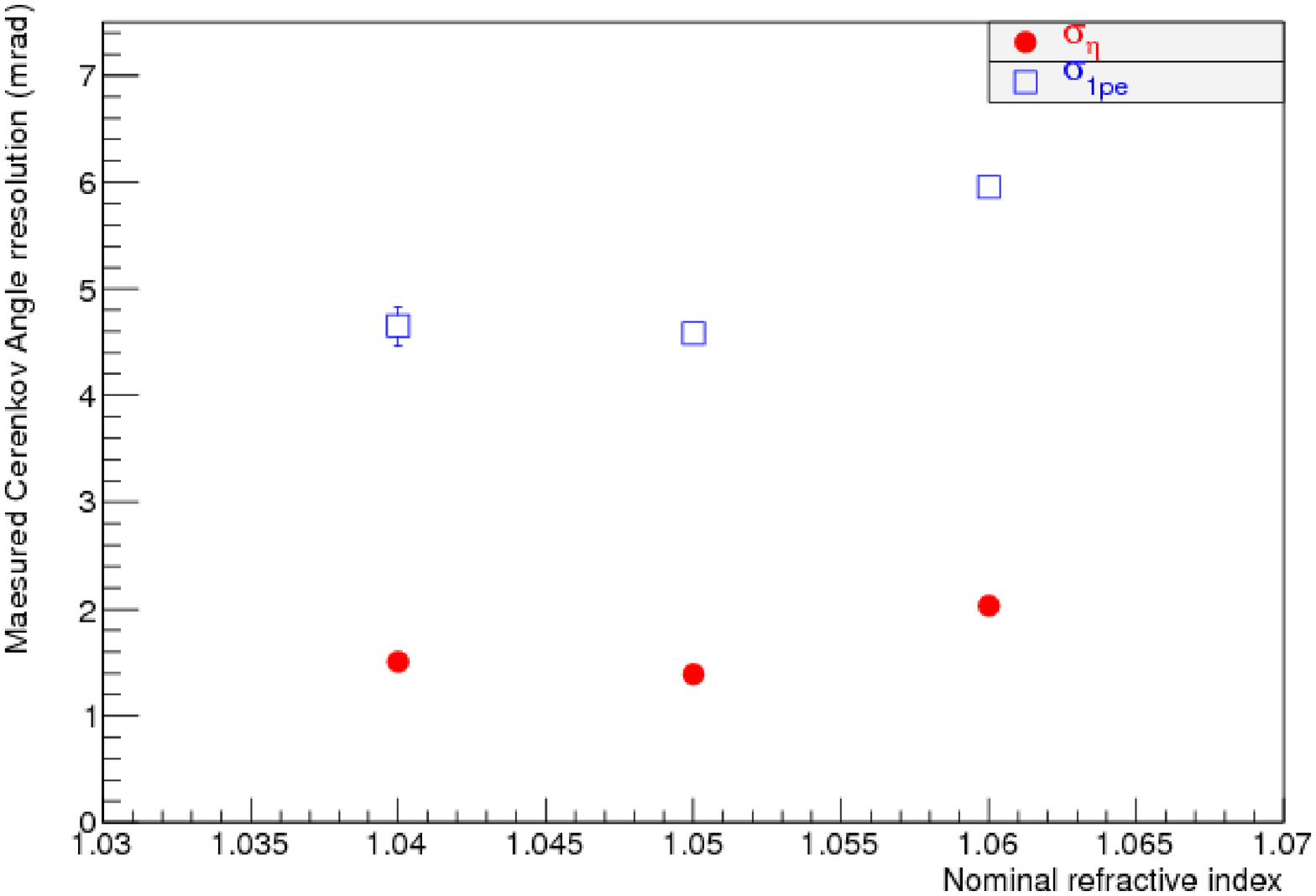}
}
\caption{Top plot: Average number of p.e. of the data (full circles) and rescaled to full ring coverage (empty squares) and expected $N_{pe}$ (dashed line). Bottom plot: Cherenkov angle resolution (full red circles) and single photon resolution $\sigma_{1pe}$ (empty blue squares).
Data for pions with 8 GeV/c momentum for aerogel radiators with same thickness $t_{rad}=20$ mm and three different nominal refractive indices.}
\label{fig:ResolVsRefInd}
\end{center}
\end{figure}

In the top plot of Fig. \ref{fig:ResolVsRefInd} we show the measured $N_{pe}$ by the full circles, while the empty squares show the results rescaled by the effective ring coverage, which is higher for lower $n$.
The number of p.e. emitted at a given wavelength and after a radiator thickness $x$ can be computed by folding the Cherenkov emission spectrum of eq. (\ref{eq:CerSpectrum}) with experimental effects, as
\begin{equation}
\label{eq:NpeCalc}
\frac{d^2N}{dxd\lambda} = \left( \frac{d^2N_C}{dxd\lambda} \right) QE(\lambda) \epsilon_{pe} e^{-\frac{t_{rad} - x}{\Lambda(\lambda) \cos(\eta_C)}}
\end{equation}
where $QE(\lambda)$ is the MAPMT quantum efficiency \cite{HamamatsuH8500}, $\epsilon_{pe} = \epsilon_{coll} \epsilon_{pack} \epsilon_{rm{SPE}}$ is the total p.e. detection efficiency and $\Lambda(\lambda)$ is the aerogel transmission length, obtained from the measured transparency $T(\lambda)$ using the relation
\begin{equation}
T(\lambda) = e^{-\frac{t_{rad}}{\Lambda(\lambda)}}.
\end{equation}
Integrating equation (\ref{eq:NpeCalc}) over $\lambda$ and $x$, we obtained the total number of p.e. values shown as the blue dashed curve in the top plot of Fig. \ref{fig:ResolVsRefInd}, which is in sufficiently good agreement with the measurements for the refractive index $n=1.04, 1.05$.
The measurement is significantly lower than the expectation for $n=1.06$.
This may indicate for this aerogel tile larger contributions from effects not considered in eq. (\ref{eq:NpeCalc}) (for example surface quality or density variations).

In the bottom plot of Fig. \ref{fig:ResolVsRefInd}, the full red circles show the measured Cherenkov angle resolution $\sigma_{\eta}$, while the empty blue squares show the single photon resolution $\sigma_{1pe}$ extracted from the $N_{pe}$ dependence using eq. (\ref{eq:SigmaFitFunction}).
While the results for the lower refractive index tiles are comparable, worse resolution has been found for the aerogel with $n=1.06$.
The larger value of $\sigma_{\eta}$ is the combination of the lower $N_{pe}$ and of the worse single photon resolution $\sigma_{1pe}$.

These results indicate the $n$ = 1.05 as the optimal choice for the CLAS12 RICH, because it provides the highest $\pi$/K separation power by combining the biggest number of p.e. with the best angular resolution.

\subsection{Chromatic dispersion}
\label{sec:ChrData}

The chromatic dispersion of the aerogel used in the tests has been studied by measuring the Cherenkov angle as a function of the photon wavelength.
This has been done by placing, just after the aerogel radiator, optical filters \cite{Edmund}, which select photons in a band of wavelengths.

The Cherenkov angle reconstruction basically extends the eq.~(\ref{eq:AngleRec}) to include refractions at the surfaces separating the air and the filters.
At each wavelength, the refractive index is then obtained from the Cherenkov angle through the eq. (\ref{eq:ref_ind}).

\begin{figure}
\begin{center}
\resizebox{0.45\textwidth}{!}{%
  \includegraphics{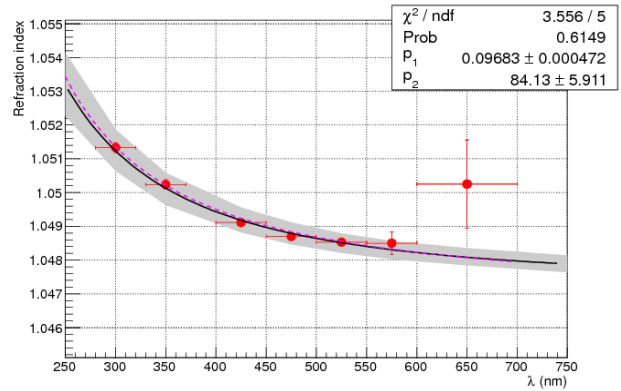}
}
\caption{Refractive index as a function of the Cherenkov light wavelength for the aerogel with $n=1.05$ and $t_{rad}=20$ mm obtained selecting pions with 8 GeV/c momentum. The full curve and the gray area represent the result of a fit with the dispersion relation eq. (\ref{eq:DispEq}), whose fitted parameters are shown in the legend, and the $1 \sigma$ error band. The dashed pink curve is the result of the calculation described in the text.}
\label{fig:RefIndVsLambda}
\end{center}
\end{figure}

In Fig. \ref{fig:RefIndVsLambda}, we show the results for the aerogel with nominal refractive index $n=1.05$ and $t_{rad}=20$ mm obtained with 8 GeV/c momentum pions.
The full circles are our data, with the vertical bars representing the 1$\sigma$ error on the mean refractive index and the horizontal ones representing the wavelength interval of each measurement.
The curve represents a fit using the Sellmeier parametrization \cite{Ghosch} of the wavelength dependence
\begin{equation}
\label{eq:DispEq}
n^2(\lambda) = 1 + \frac{p_1 \lambda^2}{\lambda^2 - p_2^2}
\end{equation}
that reproduces the data well.
The gray area in the plot represents the $1 \sigma$ error band of the fit.
From this fit, we obtained a refractive index $n_{fit}(400 \mbox{ nm}) = 1.0494 \pm 0.0004$ nm at the wavelength $\lambda_0=400$ nm.
The variation in the wavelength range between 300 and 600 nm is about $\pm 0.2 \%$ with respect to the value at 400 nm.

According to the quotation provided by the Novosibirsk factory, the refractive index at the wavelength $\lambda_0$ can be obtained by using the equation
\begin{equation}
n^2(400 \mbox{ nm}) = 1 + 0.438 \rho.
\end{equation}
Using in the measured aerogel density $\rho = 0.230$ g/cm$^3$, we got the value $n_{quoted} = 1.0492$, in agreement with our measurements.

Following Vorobiev and collaborators \cite{Vorobiev}, the $\lambda$ dependence of the aerogel refractive index can be computed from the well known values of its main components, the air and quartz, through the relation
\begin{equation}
\label{eq:RefIndTheo}
n(\lambda) = A n_{air}(\lambda) + (1 - A) n_{quartz}(\lambda).
\end{equation}
The normalization constant $A$ can be computed, if the aerogel refractive index is known at the wavelength $\lambda_0$, as
\begin{equation}
\label{eq:RefIndNorm}
A = \frac{n_{quartz}(\lambda_0) - n(\lambda_0)}{n_{quartz}(\lambda_0) - n_{air}(\lambda_0)}.
\end{equation}
Using $\lambda_0 = 400$ nm and $n(\lambda_0) = n_{fit}(400 \mbox{ nm})$, we obtained the result shown in Fig. \ref{fig:RefIndVsLambda} by the dashed pink line.
It reproduced the measured wavelength dependence very well.


\section{Analysis of the Cherenkov angle resolution}

The resolution of the measured emission angle of the Cherenkov photon can be written as
\begin{equation}
\sigma_{1pe}^2 =  \sigma_{c}^2 + \sigma_{d}^2 + \sigma_{e}^2
\end{equation}
with main contributions arising from the chromatic dispersion ($\sigma_{c}$) and from the uncertainties on the impact point measurement ($\sigma_{d}$) and on the emission point ($\sigma_{e}$).
We neglect here other contributions like background photons (largely removed in the analysis) or reconstruction issues (following the discussion on the $N_{pe}$ dependence of the measured $\rm{SPE}$ angular resolution).
We also neglect the potentially important contributions associated to the aerogel optical properties (Rayleigh scattering, surface effects, etc.) that are more difficult to determine and which require specific experimental studies.

The chromatic term is due to the variation of the aerogel refractive index with the photon wavelength and it can be further decomposed in a contribution ($\sigma_{c}^i$) due to the photon emission angle and a contribution ($\sigma_{c}^r$) due to the refraction at the crossing of the aerogel surface and can be written as
\begin{equation}
\sigma_{c}^2 =  (\sigma_{c}^i)^2 + (\sigma_{c}^r)^2 \approx \left[ \left(\frac{\partial \eta_C}{\partial n}\right)^2 + \left(\frac{\partial \theta}{\partial n}\right)^2 \right] \sigma_n^2
\end{equation}
where the variation of the air refractive index has been neglected.
The variation of the aerogel refractive index $\sigma_n$ is calculated by using the fit to our measurements, as shown in Fig. \ref{fig:RefIndVsLambda}, weighted with the spectrum of the detected photons computed by integrating eq. (\ref{eq:NpeCalc}) over the radiator thickness.

The contribution due to the uncertainty on the impact point can be written as
\begin{equation}
\sigma_{d} = \left( \frac{\partial \eta_C}{\partial R} \right) \sigma_R
\end{equation}
where, assuming a uniform photon hit probability within the square MAPMT pixels, the Cherenkov hit radius resolution is $\sigma_R = d / \sqrt{12}$, $d=6$ mm being the MAPMT pixel size.

The contribution due to the uncertainty on the emission point $x$ can be written as
\begin{equation}
\sigma_{e} = \left( \frac{\partial \eta_C}{\partial x} \right) \sigma_x
\end{equation}
where it was conservatively assumed that, on average, the photon has been emitted at half of the aerogel radiator thickness, thus $ \sigma_x \approx t_{rad} / \sqrt{12}$. 

\begin{figure}
\begin{center}
\resizebox{0.35\textwidth}{!}{%
  \includegraphics{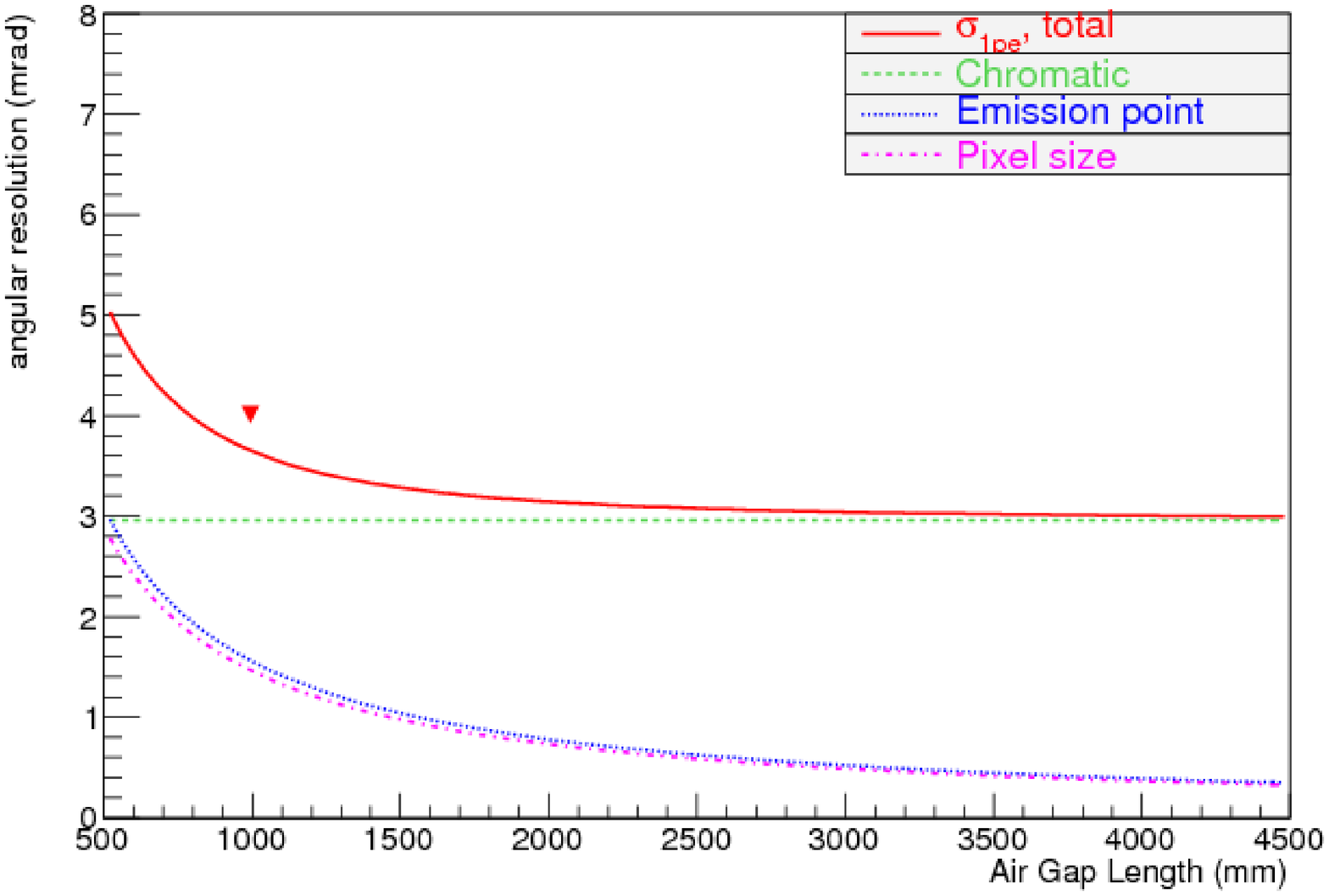}
}
\resizebox{0.35\textwidth}{!}{%
  \includegraphics{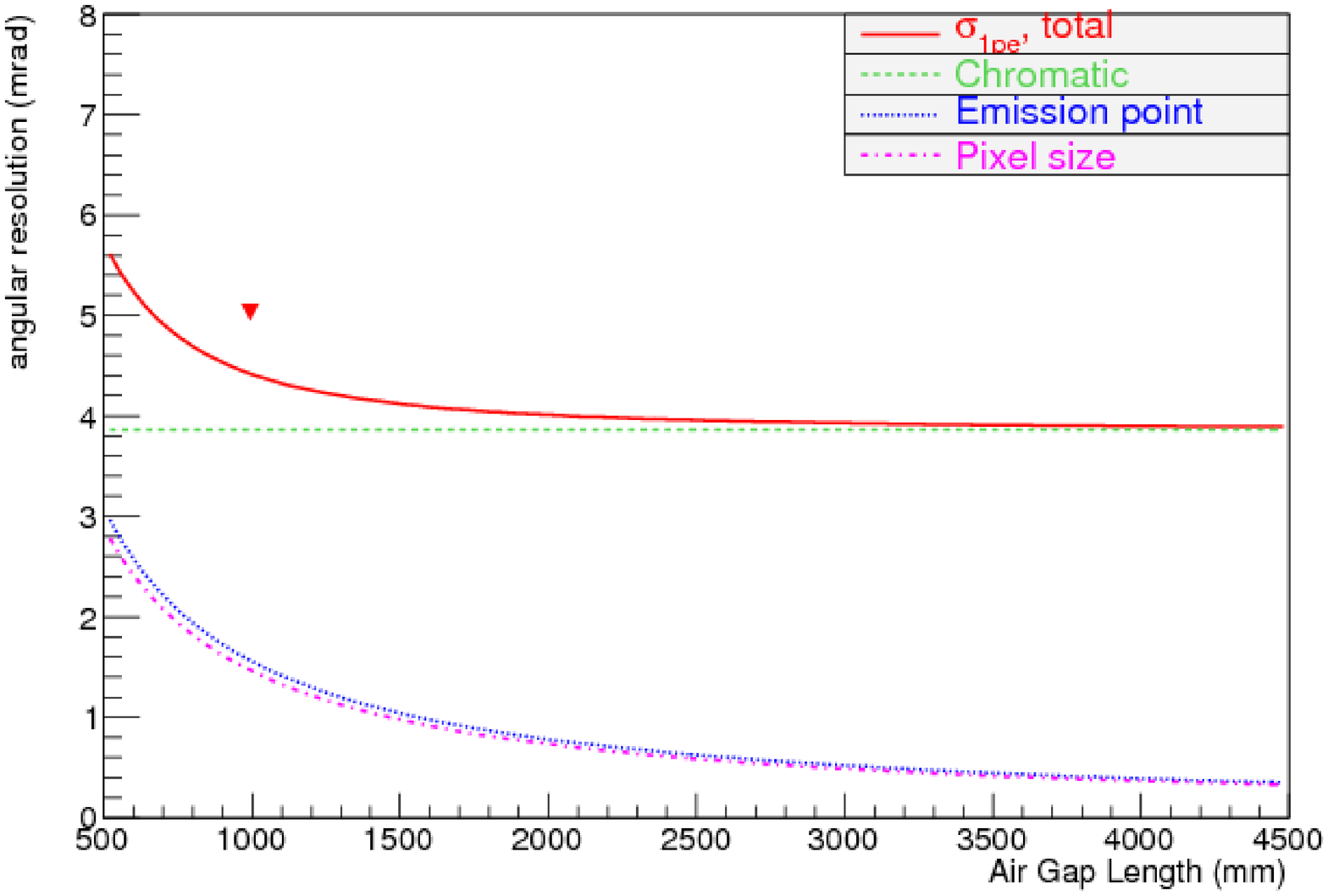}
}
\caption{Calculated contributions to the Cherenkov photon angular resolution $\sigma_{1pe}$ as a function of the air gap length for the fourteen H8500C with normal glass window (top plot) and for the fourteen H8500C-03 with UV-enhanced glass window (bottom plot): the curves represent the calculated resolutions, while the points represent our measurement. Data for 8 GeV/c pions and aerogel with $n=1.05$ and $t_{rad}=20$ mm.}
\label{fig:CalcResVsPmt}
\end{center}
\end{figure}

In Fig. \ref{fig:CalcResVsPmt}, we show the result of the calculated $\sigma_{1pe}$ for the 14 H8500C with normal glass window (top plot) and for the 14 H8500C-03 with UV-enhanced glass window (bottom plot).
The various curves are plotted as a function of the gap length, for an aerogel radiator thickness $t_{rad}= 20$ mm.
The emission point and the pixel size contributions have similar size, decreasing with the gap length, and give a significant contribution only for a gap length of the order of 1 m or less.
They are also independent of the MAPMT type.
The chromatic term is almost flat and is the dominant one, except for small gap length below approximately $1$ m.

We also show in the figure the measured total $\sigma_{1pe}$ values at the gap length $t_{air} = 994$ mm of our prototype.
The calculation results for the H8500C MAPMTs agree with the measurements within less than $10\%$.
The total of possible missing contributions can be at most of the same order of magnitude of the smaller ones in the top plot of Fig. \ref{fig:CalcResVsPmt}.
For the H8500C-03 data, the discrepancy is slightly larger, even though below $15\%$, indicating that the missing contributions (as for example the Rayleigh term) might be larger in this case.
The almost $30\%$ increase in the chromatic term due to the larger wavelength interval extending into the UV region is anyway not compensated by the about $20\%$ increase in the photon flux measured for the H8500C-03 MAPMTs. 
This supports the results of Sect. \ref{sec:UVData} that the UV photons may worsen the resolution in spite of increasing the number of photoelectrons.

\section{Conclusions}

A RICH detector has been proposed to extend the PID capabilities of the CLAS12 spectrometer of the Hall B at JLab in the momentum range between 3 and 8 GeV/c.
The detector presents an innovative hybrid configuration with an aerogel radiator, MAPMTs as photon detectors and mirrors, in which the Cherenkov rings are imaged either directly (for forward particles) or after two mirror reflections and a double pass through the aerogel (for large angle particles).

We reported here on the beam tests of a large scale prototype of the detector performed at the T9 hadron beam line at CERN for the proximity focusing configuration.
The setup used multianode photomultiplier tubes Hamamatsu H8500 (a device not optimized for the single photon regime) and readout electronics based on the MAROC3 chip.
The results showed that they provide good performances in the single photon regime.
The radiator was aerogel with various refractive index values.

In the most challenging conditions expected for the CLAS12 RICH, i.e. 8 GeV/c momentum pions, we measured $N_{pe} = 12.02 \pm 0.02$ p.e. per ring and $\sigma_{1pe} = 4.58 \pm 0.02$ mrad.
These results, that provided the required pion to kaon rejection factor of 1:500, have been obtained with an aerogel radiator with $n=1.05$ and $t_{rad}=20$ mm, the final choice for the CLAS12 RICH.
The measured resolutions are in reasonable agreement with the expectations.

The tests also included measurements of several aspects of the foreseen RICH detector: contribution of UV photons to the angular resolution, aerogel refractive index dependence and chromatic dispersion.
These measurement showed that all these effects are effectively under control, without affecting the required performances of the CLAS12 RICH.

%

\begin{thebibliography}{}
%
%

\bibitem{CLAS12} CLAS12 Technical Design Report, version 5.1, 208 (2008).
\bibitem{CLAS12_phys} V.D. Burkert {\it et. al.}, arXiv:1203.2373 [hep-ex] (2012).
\bibitem{PSHP} H. Avakian {\it et. al.}, arXiv:1202.1910v2 [hep-ex] (2012).

\bibitem{RICH_sim1} M. Contalbrigo {\it et al.}, Nucl. Instrum. Meth. \textbf{ A 639}, 302 (2011).
\bibitem{RICH_sim2} A. El Alaoui {\it et al.}, Physics Procedia \textbf{ 37}, 773 (2012).

\bibitem{Cont2014} M. Contalbrigo {\it et. al.}, Nucl. Instrum. Meth. \textbf{ A 766}, 22 (2014).

\bibitem{BTF} G. Mazzitelli {\it et al.}, Nucl. Instrum. Meth. \textbf{ A 515}, 524 (2003).
\bibitem{CERN_T9} http://sba.web.cern.ch/sba/Documentations/Eastdocs/\\/docs/T9\_Guide.pdf
\bibitem{CERN_T9_optics} L. Durieu, M. Martini and A.-S. Muller, Proceedings of the 2001 Particle Accelerator Conference, Chicago.

\bibitem{Aerogel_old1} R. De Leo {\it et al.}, Nucl. Instrum. Meth. \textbf{ A 595}, 19 (2008).
\bibitem{Aerogel_old2} A. Yu. Barnyakov {\it et al.}, Nucl. Instrum. Meth. \textbf{ A 453}, 326 (2000). 
\bibitem{Aerogel_old3} R. Pereira {\it et al.}, Nucl. Instrum. Meth. \textbf{ A 639}, 37 (2011).
\bibitem{Aerogel_old4} R. Forty {\it et al.}, Nucl. Instrum. Meth. \textbf{ A 623}, 294 (2010).

\bibitem{Aerogel_new} T. Iijima {\it et al.}, Nucl. Instrum. Meth. \textbf{ A 598}, 138 (2009).

\bibitem{BINP1} A. Yu. Barnyakov {\it et. al.}, Nucl. Instrum. Meth. \textbf{ A 639}, 225 (2011).
\bibitem{BINP2} A. Yu. Barnyakov {\it et. al.}, Nucl. Instrum. Meth. \textbf{ A 732}, 352 (2014).

\bibitem{AerogelTests} L. L. Pappalardo, EPJ Web Conf. \textbf{ 73}, 08003 (2014).

\bibitem{SnellDescartes} T. Bellunato {\it et. al.}, Eur. Phys. J. \textbf{ C 52}, 759 (2007).


\bibitem{HamamatsuH8500} http://www.hamamatsu.com/resources/pdf/etd/\\/H8500\_H10966\_TPMH1327E.pdf

\bibitem{hamamatsu_pmt} Photomultiplier tubes, https://www.hamamatsu.com/\\/resources/pdf/etd/PMT\_handbook\_v3aE.pdf.

\bibitem{MAPMT_LNF} R. A. Montgomery {\it et. al.}, Nucl.Instrum.Meth. \textbf{ A 790}, 28 (2015).

\bibitem{MAPMT_Glasgow} R. A. Montgomery {\it et al.}, Nucl. Instrum. Meth. \textbf{ A 695}, 326 (2012).

\bibitem{GEM} F. Sauli, Nucl. Inst. Methods \textbf{ A 386}, 531 (1997).

\bibitem{CERN_GEM} CERN GEM production web page: http://ts-dep-dem.web.cern.ch/ts-dep-dem/products/gem/.

\bibitem{MAROC} S. Blin {\it et al.}, MAROC3 datasheet, October 2010, OMEGA website: http://omega.in2p3.fr.

\bibitem{Argentieri} A.G. Argentieri {it et al.}, Nucl. Instrum. Meth. \textbf{ A 617}, 348 (2010).

\bibitem{Bellini} V. Bellini {\it et al.}, 2012 JINST 7 C05013.

\bibitem{APV25} M.J. French {\it et al.}, Nucl. Instrum. Meth. \textbf{ A 466}, 359 (2002).

\bibitem{minuit} http://seal.web.cern.ch/seal/snapshot/work-packages/mathlibs/minuit/.
\bibitem{ALICE} ALICE Technical Design Report, CERN/LHCC 98-19.

\bibitem{JLab_HallA} G.M. Urciuoli {\it et al.}, Nucl. Instrum. Meth. \textbf{ A 612}, 56 (2009).

\bibitem{Edmund} http://www.edmundoptics.eu/optics/optical-filters/.

\bibitem{Ghosch} G. Ghosch, Appl. Opt. \textbf{ Vol. 36}, 1540 (1997).

\bibitem{Vorobiev} V.I. Vorobiev {\it et. al}, Proceedings of the Workshop on Physics and Detection for DA$\Phi$NE. Report-INFN-Frascati. 1991.


\end{thebibliography}
%

\end{document}